\documentclass[10pt,journal,twoside]{IEEEtran}

\usepackage[english]{babel}
\usepackage{graphicx}
\usepackage{subfigure}
\usepackage{multicol}
\usepackage{setspace}
\usepackage{amsmath}
\usepackage{amsthm}
\usepackage{epstopdf}
\usepackage{fancyhdr}
\usepackage{indentfirst}
\usepackage{enumerate}
\usepackage{caption}
\usepackage{leftidx}
\usepackage{amssymb}
\usepackage{float}
\usepackage{breqn}
\usepackage{color, xcolor}
\usepackage{bm}
\usepackage{cite}
\usepackage{multirow}
\usepackage{totcount}
\usepackage{algorithm}
\usepackage{algcompatible}
\usepackage{hyperref}
\usepackage{bbm}
\usepackage{dutchcal}
\usepackage{chngcntr}
\usepackage{setspace}
\counterwithin*{footnote}{page}

\newtheoremstyle{colon}%
{}
{}
{\itshape}
{}
{\itshape}
{:}
{ }
{\thmname{#1}\ \!\thmnumber{\itshape#2}\thmnote{(#3)}}

\theoremstyle{colon}
\newtheorem*{Remark*}{Remark}
\newtheorem*{Theorem*}{Theorem}

\newtheorem*{Lemma*}{Lemma}

\newtheorem{Prop}{Proposition}
\numberwithin{Prop}{section}

\newtheorem{Def}{Definition}
\numberwithin{Def}{section}

\newtheorem{Rem}{Remark}
\numberwithin{Rem}{section}

\newtheorem{Thm}{Definition}
\numberwithin{Thm}{section}

\DeclareMathOperator*{\argmin}{arg\,min}
\DeclareMathOperator*{\argmax}{arg\,max}

\newcommand{\ts}{\textsuperscript}

\begin{document}
\title{Coexistence between Task- and Data-Oriented Communications: A Whittle's Index Guided Multi-Agent Reinforcement Learning Approach}
\author{Ran Li, Chuan Huang, Xiaoqi Qin, Shengpei Jiang, Nan Ma, and Shuguang Cui
\thanks{This work was submitted in part to 2022 IEEE Global Communications Conference.

R. Li is with the School of Science and Engineering (SSE) and the Future Network of Intelligence Institute (FNii), The Chinese University of Hong Kong, Shenzhen 518172, China, (e-mail: ranli2@link.cuhk.edu.cn).

C. Huang and S. Cui are with the School of Science and Engineering (SSE) and Future Network of Intelligence Institute (FNii), The Chinese University of Hong Kong, Shenzhen 518172, China, and with Peng Cheng Laboratory, Shenzhen 518066, China, (e-mails: huangchuan@cuhk.edu.cn; shuguangcui@cuhk.edu.cn).

X. Qin is with the State Key Laboratory of Networking and Switching Technology, Beijing University of Posts and Telecommunications, Beijing 100876, China, (e-mail: xiaoqiqin@bupt.edu.cn).

S. Jiang is with the SF Technology, Shenzhen 518052, China, (e-mail: philip.jiang@sfmail.sf-express.com).

N. Ma is with the State Key Laboratory of Networking and Switching Technology, Beijing University of Posts and Telecommunications, Beijing 100876, China, and with the Department of Broadband Communication, Peng Cheng Laboratory, Shenzhen 518066, China, (e-mail: manan@bupt.edu.cn).
}
}
\maketitle
\thispagestyle{empty}
\begin{abstract}
We investigate the coexistence of task-oriented and data-oriented communications in a IoT system that shares a group of channels, and study the scheduling problem to jointly optimize the weighted age of incorrect information (AoII) and throughput, which are the performance metrics of the two types of communications, respectively. This problem is formulated as a Markov decision problem, which is difficult to solve due to the large discrete action space and the time-varying action constraints induced by the stochastic availability of channels. By exploiting the intrinsic properties of this problem and reformulating the reward function based on channel statistics, we first simplify the solution space, state space, and optimality criteria, and convert it to an equivalent Markov game, for which the large discrete action space issue is greatly relieved. Then, we propose a Whittle's index guided multi-agent proximal policy optimization (WI-MAPPO) algorithm to solve the considered game, where the embedded Whittle's index module further shrinks the action space, and the proposed offline training algorithm extends the training kernel of conventional MAPPO to address the issue of time-varying constraints. Finally, numerical results validate that the proposed algorithm significantly outperforms state-of-the-art age of information (AoI) based algorithms under scenarios with insufficient channel resources.

\end{abstract}
\section{Introduction}
In the past decades, the spotlight of Internet of things (IoT) is shifting towards enabling autonomous networked control applications that require timely status updates \cite{iot1}, e.g., environmental monitoring \cite{iot2}, emergency detection \cite{iot3}, and healthcare systems \cite{iot4}. Under such scenarios, IoT devices are deployed to obtain real-time awareness of a monitored physical process by continuous sampling and data uploading. Considering the time-varying nature of environment, the freshness of status updates is of critical importance to the performance of subsequent tasks. Age of information (AoI) \cite{aoi_original} is proposed as a performance metric for such task-oriented communications, which is defined as the time elapsed since the latest received status update was generated at the monitoring device. Note that conventional resource provisioning schemes follow the data-oriented design philosophy of maximizing network-level throughput. Therefore, in order to accommodate time-sensitive data transmission in task-oriented communications, one straightforward approach is to reserve spectrum for devices with critical status updates to guarantee the timely delivery of data. However, such separate design degrades the spectrum efficiency due to the intermittent nature of monitoring devices. Therefore, it is important to establish a flexible coexistence strategy between task-oriented and data-oriented communications. This problem, if left unsolved, will jeopardize the utility of networked control systems.

The focus of recent research on resource scheduling strategies for time-sensitive applications is optimizing AoI among multiple devices. The authors in \cite{AoI_03} studied the scenario with multiple IoT devices monitoring multiple processes and then transmitting their statuses to one central base station (BS) over one channel, and a near-optimal algorithm to minimize the average AoI was proposed. The authors in \cite{AoI_04} extended the previous work to the multiple channels case, and proposed a low-complexity algorithm based on the Lagrangian relaxation method \cite{whittle}. The authors in \cite{AoI_01} studied the AoI minimization problem in a time-framed system, where multiple statuses can be transmitted during one time frame, and compared the performances among the randomized policy, the max-weight policy \cite{neely}, and the Whittle's index algorithm. In order to strike a balance between the achievable throughput and age under sparse sampling at source nodes, various strategies have been proposed \cite{joint_1,joint_2,joint_3}. The authors in \cite{joint_1} studied the scenario with multiple IoT devices, and aimed to minimize their average AoIs while simultaneously satisfy the constraints on their throughputs. In \cite{joint_2}, IoT was utilized to simultaneously monitor one process with a single monitoring device and collect data from multiple traditional devices, and the average AoI at the monitoring device and the throughputs of the traditional devices were evaluated under the ALOHA protocol. The authors in \cite{joint_3} considered the most general scenario with multiple monitoring devices acquiring statuses from multiple processes and multiple traditional devices collecting data, and a Lyapunov drift based scheduling policy was proposed to jointly optimize the average AoI and throughput.

Recently, age of incorrect information (AoII) \cite{AoII_original_1} is proposed as a more advanced performance metric that captures not only the aging of status updates, but more importantly the change of context of the monitored process at source node. To elaborate, the authors in \cite{AoII_original_1,AoII_original_2} discussed the scenario where a base station (BS) in IoT predicts the state of a process. Apparently, the BS prefers to collect new status information from the monitoring device when the state of the considered process indeed changes, not when the currently reserved status at the BS is not fresh. This indicates that AoII is more compatible with the prediction tasks than AoI. 

There has been growing research interest in designing AoII-based scheduling policy for monitoring devices \cite{AoII_infocom_binary,AoII_chen1_binary,AoII_chen2,AoII_saad1,AoII_original_1}. Specifically, the authors in \cite{AoII_infocom_binary,AoII_chen1_binary} discussed the scenario where the target process to be predicted simply follows the binary distribution, and the optimal policy for the case with one target process and a low-complexity suboptimal policy for the case with multiple target processes were developed. One step further, the authors in \cite{AoII_chen2,AoII_saad1,AoII_original_1} discussed a more complex prediction scenarios, where the target processes have more than two states. Specifically, the authors in \cite{AoII_chen2,AoII_saad1} discussed the scenario where the state transitions of the processes only occur between the adjacent states, and a Whittle's index based suboptimal policies were proposed for the case with multiple target processes. The author in \cite{AoII_original_1} considered the scenario where the state transitions between any two states are allowable, and derived the optimal policy for the case with one target process. However, the optimization framework in \cite{AoII_original_1} cannot be directly extended to the case with multiple target processes and no existing work has touched this case. 

In this paper, we employ AoII as the performance metric for task-oriented communications and investigate the coexistence strategy of task-oriented and data-oriented communications by jointly optimizing the average AoII and the throughput. Specifically, we study a general scenario: multiple process-aware monitoring devices monitor and transmit the status updates of multiple random processes, respectively, which have more than two states and the state transition between any two states is allowable; multiple traditional devices purse high throughput and each round of the data transmissions for the traditional devices may last for multiple time slots; limited channel resources are available in IoT for the data transmissions of the monitoring and traditional devices. We summarize our contributions as follows:
\begin{itemize}
	\item We formulate the joint optimization problem as a Markov decision problem, which is challenging due to the large solution space, large state space, and average optimality criteria. Moreover, it has a large discrete action space and time-varying action constraints induced by the stochastic availability of channels, where existing algorithms to solve Markov decision problems cannot efficiently address the problems of this type. To overcome these challenges, we first analyze the intrinsic properties of this Markov decision problem, and prove that there exist stationary policies to achieve its optimum. Next, based on this stationary feature, we reformulate the reward function and transform the original problem to an equivalent form with a much smaller state space and a simplified state transitions. Then, we validate the existence of the Blackwell policies for the equivalent problem, replace the average optimality criteria by the discounted version, and prove that the optimal policies under this discounted criteria also optimize the problem with average optimality criteria. Finally, we convert the above discounted Markov decision problem as an equivalent Markov game by treating each channel as individual agent, for which the large discrete action space issue is relieved. Remarkably, all these problem simplifications and conversions are theoretically validated to be equivalent.
	\item We propose a multi-agent reinforcement learning algorithm, namely the Whittle's index guided multi-agent proximal policy optimization (WI-MAPPO), to efficiently solve the proposed Markov game. Specifically, WI-MAPPO deploys a Whittle's index guided action fusion module to further shrink the action space of the Markov game. To design this module, we first prove that Whittle's index for the monitoring devices exists by validating their indexabilities, and then utilize an efficient exhausted searching algorithm to approximate the Whittle's index. Finally, we construct this module by generating a sufficient large Whittle's index table. Moreover, we modify both the actor network and the probability ratio derivation of the training algorithm for multi-agent proximal policy optimization (MAPPO) to train the proposed WI-MAPPO. By doing so, the training procedure will not violate the time-varying constraints and ensure even faster and more accurate estimations on the advantage functions of MAPPO. Remarkably, although the time-varying constraints issue shrinks the solution space of the considered Markov game and is doomed to induce loss of optimality compared with the non-constrained Markov game, it is validated that the proposed WI-MAPPO can greatly narrow this optimality gap and achieve almost the same performances for the constrained and non-constrained Markov games.
\end{itemize}

The remainder of this paper is organized as follows. Section \ref{Section2} introduces the system model and formulates the joint scheduling problem. Section \ref{Section3} presents the proposed algorithm. Section \ref{Section4} evaluates the performance of the proposed algorithm. Finally, Section \ref{Section5} concludes this paper.

\begin{figure*}[!htb]
\centering
\includegraphics[width=6.8in]{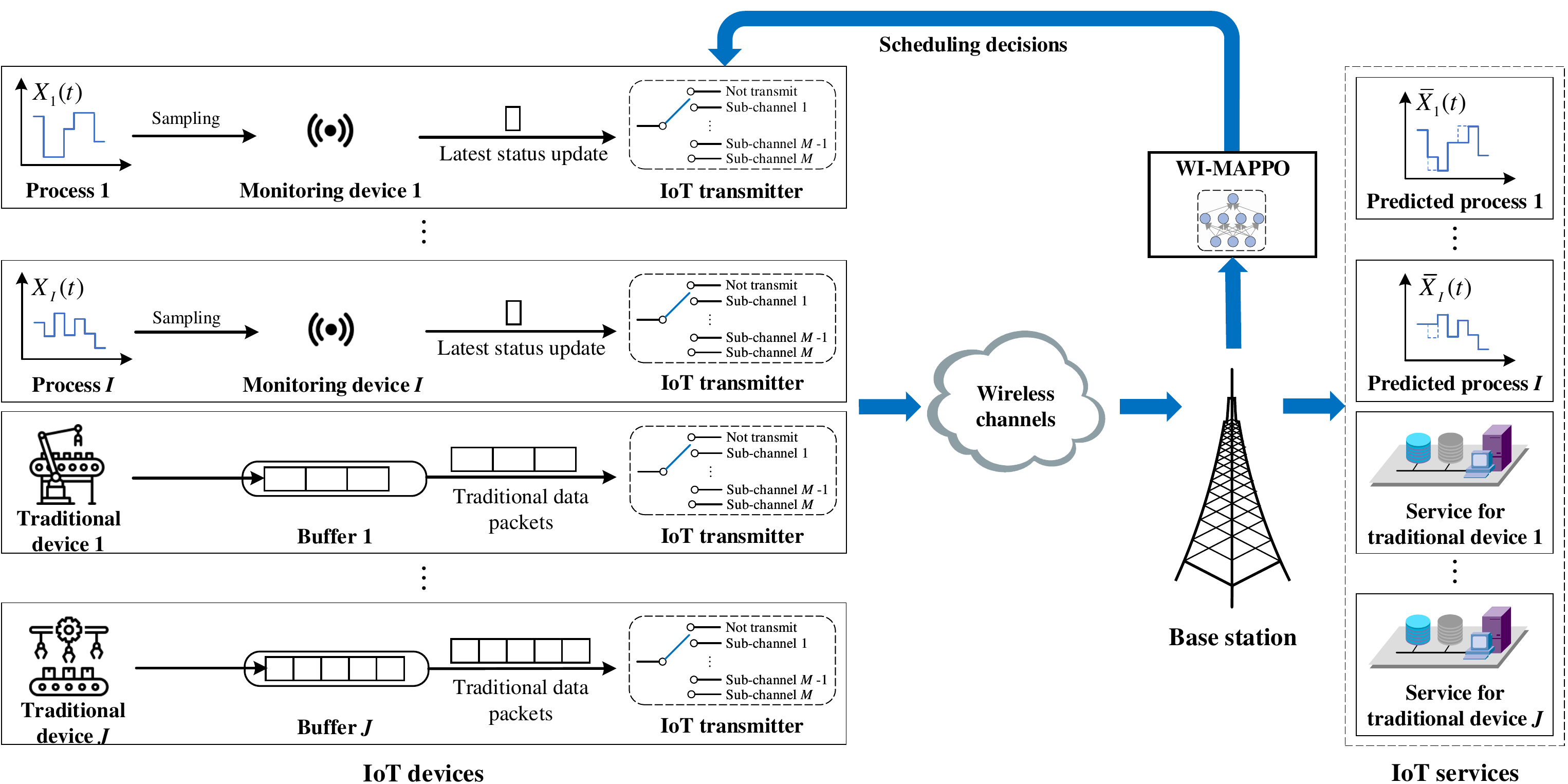}
\caption{Joint schedule of the task-oriented and data-oriented communications.}\label{fig_system}
\end{figure*}

\section{System Model and Problem Formulation}\label{Section2}
Consider a slotted IoT system as shown in Fig.~\ref{fig_system}, which consists of a BS and a set of IoT devices. The set of devices consists of two types, including $I$ monitoring devices to enable real-time situational awareness at BS and $J$ traditional devices for data collection. The monitoring devices are deployed at $I$ different monitoring spots to obtain real-time perceptions about $I$ different random process $\{X_i(t)\}_{i=1}^I$ by periodical data sampling and uploading, and the scheduling performance for monitoring devices is quantified by AoII, which captures the semantic attributes of information in terms of the relevance of transmitted data to the subsequent task. As for traditional devices, the scheduling performance can be quantified by the achievable network-level throughput. We assume an OFDMA-based system with $M$ sub-channels. The data transmission for traditional devices may last for multiple consecutive time slots depending on the size of sampled packets, while the transmission for monitoring devices is assumed to be completed within one time slot due to small packet size for status update information. Notably, we also call the $i$\ts{th} monitoring device as the $i$\ts{th} device and the $j$\ts{th} traditional device as the $(I+j)$\ts{th} device.

\subsection{System model}
The joint scheduling between the task-oriented and data-oriented communications is formulated as a Markov decision problem and we introduce its state, action, transitions, and reward as follows.
\subsubsection{State} The state contains {\it AoII vector}, {\it channel gains}, and {\it channel availability}.

\textbf{AoII vector}: AoII counts the age of the incorrect prediction and increases as long as the predicted state is incorrect. Particularly, it is defined as follows \cite{AoII_original_1}.
\begin{Def}[AoII]
The AoII at the $i$\ts{th} monitoring device in the $t$\ts{th} time slot is denoted as $x_i(t)\in\mathbb{Z}_{\geq0}$, where $\mathbb{Z}_{\geq0}$ is the set of all non-negative integers, and recursively defined by
\begin{align}\label{def:x}
x_i(t+1)\triangleq\left\{\begin{array}{ll}
0&\bar{X}_i(t+1)=X_i(t+1)\\
x_i(t)+1&\bar{X}_i(t+1)\neq X_i(t+1).
\end{array}\right.
\end{align}
where $\bar{X}_i(t)$ is the prediction on $X_i(t)$ made by the BS. 
\end{Def}
Then, the {\it AoII vector} at $I$ monitoring devices is defined as $\bm{x}(t)\triangleq[x_1(t),x_2(t),\cdots,x_I(t)]^T$.

\textbf{Channel gains:} Denote the channel coefficient and the channel gain of the link between the BS and the $j$\ts{th} traditional device over the $m$\ts{th} channel at the $t$\ts{th} time slot as $h_{j,m}(t)$ and $g_{j,m}(t)$, respectively, i.e., $g_{j,m}(t)=|h_{j,m}(t)|^2$. Then, define the {\it channel gains} at the $t$\ts{th} time slot as matrix $\bm{G}(t)\in\mathcal{G}^{J\times M}$, where the $(j,m)$\ts{th} entry of $\bm{G}(t)$ is $g_{j,m}(t)$, i.e., $[\bm{G}(t)]_{(j,m)}\triangleq g_{j,m}(t)$, and $\mathcal{G}$ is the value space of $g_{j,m}(t)$ and considered as a finite set.

\textbf{Channel availability:} Each data transmission for the $j$\ts{th} traditional device is considered to consume $T_j\in\mathbb{Z}^{+}$ consecutive time slots with $\mathbb{Z}^{+}$ being the set of all positive integers, during which the occupied channel is not available for new data transmission. Specifically, denote the availability condition of the $m$\ts{th} channel for the data transmission of the $j$\ts{th} traditional device at the $t$\ts{th} time slot as $b_{j,m}(t)\in\{0,1,\cdots,T_j-1\}$: if the $m$\ts{th} channel is not transmitting data for the $j$\ts{th} traditional device at the $t$\ts{th} time slot, $b_{j,m}(t)$ is set to 0; otherwise, $b_{j,m}(t)$ is equal to the number of the remaining time slots for the release of the $m$\ts{th} channel. Then, define the {\it channel availability} as a $J$-by-$M$-dimension matrix $\bm{B}(t)$, where the $(j,m)$\ts{th} entry of $\bm{B}(t)$ equals $b_{j,m}(t)$, i.e., $[\bm{B}(t)]_{(j,m)}\triangleq b_{j,m}(t)$. Note that the status update transmission of monitoring devices is assumed to be completed within one time slot, and thus the allocated channel will always be released for new data transmission at the next time slot. 

Denote the state of the considered Markov decision problem at the $t$\ts{th} time slot as $\bm{s}(t)$. Obviously, $\bm{s}(t)=(\bm{x}(t),\bm{G}(t),\bm{B}(t))$ holds and the state space $\mathcal{S}=\mathbb{Z}_{\geq0}^{I\times1}\times\mathcal{G}^{J\times M}\times\mathcal{B}$ is countable, where $\mathcal{B}$ is the value space of $\bm{B}(t)$.

\subsubsection{Action}\label{action_setting}
Denote the scheduling decision for the $m$\ts{th} channel at the $t$\ts{th} time slot as $a_m(t)\in\{0,1,\cdots,I+J\}$. Specifically, $a_m(t)=0$ means that the $m$\ts{th} channel is not scheduled to start a new data transmission; and $a_m(t)>0$ means to transmit data for the $a_m(t)$\ts{th} device over the $m$\ts{th} channel. Then, denote the scheduling decision for all the channels at the $t$\ts{th} time slot as $\bm{a}(t)\triangleq[a_1(t),a_2(t),\cdots,a_M(t)]^T$, and apparently $\bm{a}(t)$ is the action of the considered Markov decision problem. 

Remarkably, the actions in $\{0,1,\cdots,I+J\}^{M\times 1}$ are not always allowable. If the $m$\ts{th} channel has been reserved for data transmission in previous slots, i.e., $\sum_{j=1}^Jb_{j,m}(t)>0$ holds, it cannot be scheduled for any new data transmission. That is, the action is constrained by
\begin{align}\label{con:P}
	\sum_{j=1}^Jb_{j,m}(t)a_m(t)=0,\ \forall\ m\in\{1,2,\cdots,M\}.
\end{align}
Then, we denote the allowable action space at state $\bm{s}(t)$, which contains all the actions satisfying the constraints in \eqref{con:P}, as $\mathcal{A}_{\bm{s}(t)}$.

\subsubsection{Transitions} The transitions are to update {\it AoII vector}, {\it channel gains}, and {\it channel availability}.

\textbf{AoII vector:} The transitions of AoII vector needs the knowledge of the transitions of $\{X_i(t)\}_{i=1}^I$. Particularly, this paper considers that the state space of $\{X_i(t)\}$, denoted as $\mathcal{X}_i$, contains $|\mathcal{X}_i|$ real numbers, the self-transition probability of $\{X_i(t)\}$, denoted as $\text{Pr}\{X_i(t+1)=x|X_i(t)=x\}$ with $x$ being any state in $\mathcal{X}_i$, equals $p_i$, and the probability of transition to any other state, denoted as $\text{Pr}\{X_i(t+1)=x|X_i(t)=y\}$ with $x\neq y$ and $y$ being any other state in $\mathcal{X}_i$, equals $q_i\triangleq\frac{1-p_i}{|\mathcal{X}_i|-1}$.

Now, we study the transitions of $x_i(t)$ and first consider the case that we choose to transmit data for the $i$\ts{th} monitoring device over some channel at the $t$\ts{th} time slot, i.e., $\sum_{m=1}^{M}\mathbbm{1}_i\left(a_m(t)\right)>0$, where the indicator function $\mathbbm{1}_i(x)$ equals 1 if $x$ equals $i$ and otherwise, it equals 0. For this case, state $X_i(t)$ is transmitted to the BS during the $t$\ts{th} time slot and the BS makes the prediction for the $(t+1)$\ts{th} time slot by $\bar{X}_i(t+1)=X_i(t)$. Apparently, $\bar{X}_i(t+1)=X_i(t+1)$ holds with probability $p_i$ since $X_i(t+1)=X_i(t)$ holds with probability $p_i$. Therefore, based on \eqref{def:x}, when $\sum\nolimits_{m=1}^{M}\mathbbm{1}_i\left(a_m(t)\right)>0$, we have
\begin{align}\label{trans:x1}
x_i(t+1)=\left\{\begin{array}{ll}
	0&\text{with probability }p_i\\
	x_i(t)+1&\text{with probability }1-p_i.
\end{array}\right.
\end{align}

Then, we consider the other case that we choose not to transmit data for the $i$\ts{th} monitoring device over any channel, i.e., $\sum_{m=1}^{M}\mathbbm{1}_i\left(a_m(t)\right)=0$. For this case, the BS has to inherit its previous prediction, i.e., $\bar{X}_i(t+1)=\bar{X}_i(t)$, and $x_i(t)$ updates itself based on the following rules:
\begin{itemize}
	\item When $x_i(t)=0$, $\bar{X}_i(t+1)=X_i(t+1)$ holds with probability $p_i$, since $\bar{X}_i(t+1)=\bar{X}_i(t)=X_i(t)$ holds for sure and $X_i(t)=X_i(t+1)$ holds with probability $p_i$. Then, based on \eqref{def:x}, when $\sum_{m=1}^{M}\mathbbm{1}_i\left(a_m(t)\right)=0$ and $x_i(t)=0$, we have
	\begin{align}\label{trans:x2}
        x_i(t+1)\!=\!\left\{\begin{array}{ll}
        \!0&\text{with probability }p_i\\
        \!x_i(t)+1&\text{with probability }1-p_i,
        \end{array}\right.
    \end{align}
	\item When $x_i(t)>0$, $\bar{X}_i(t+1)=X_i(t+1)$ holds with probability $q_i$, since $\bar{X}_i(t+1)=\bar{X}_i(t)$ holds for sure and $\bar{X}_i(t)=X_i(t+1)$ holds with probability $q_i$. Then, based on \eqref{def:x}, when $\sum_{m=1}^{M}\mathbbm{1}_i\left(a_m(t)\right)=0$ and $x_i(t)>0$, we have
	\begin{align}\label{trans:x3}
        x_i(t+1)\!=\!\left\{\begin{array}{ll}
        \!0&\text{with probability }q_i\\
        \!x_i(t)+1&\text{with probability }1-q_i.
        \end{array}\right.
    \end{align} 
\end{itemize}

\textbf{Channel gains:} For the link between the BS and any traditional device, the channel coefficient of this link is modeled as a stationary ergodic process, and so is the channel gain. Particularly,
\begin{align}\label{trans:g}
&\text{Pr}\{g_{j,m}(t+1)=g'|g_{j,m}(t)=g\}\nonumber\\
=&\text{Pr}\{g_{j,m}(1)=g'|g_{j,m}(0)=g\}\triangleq\text{Pr}_{j,m}\{g'|g\},
\end{align}
holds for all $t\in\mathbb{Z}^{+},\ j\in\{1,2,\cdots,J\},$ $m\in\{1,2,\cdots,M\}$, and $g, g'\in\mathcal{G}$, where $\text{Pr}_{j,m}\{g'|g\}$ is a constant and represents the probability for $g_{j,m}(t)$ transiting from $g$ to $g'$. We also consider $\mathcal{G}\triangleq\{g_1,g_2,\cdots,g_{|\mathcal{G}|}\}$ as a finite real number set with $0<g_1\leq g_2\leq\cdots\leq g_{|\mathcal{G}|}$.

\textbf{Channel availability:} We update the {\it channel availability} in the following four cases: If the $m$\ts{th} channel is currently transmitting data for the $j$\ts{th} traditional device, i.e., $b_{j,m}(t)>0$, the remaining time for the release of the $m$\ts{th} channel decreases by one at the next time slot, i.e., $b_{j,m}(t+1)=b_{j,m}(t)-1$; if the $m$\ts{th} channel is reserved by another traditional device, i.e., $b_{j,m}(t)=0$ and $\sum_{j'\neq j}b_{j',m}(t)>0$, $b_{j,m}(t+1)$ remains 0; if the $m$\ts{th} channel is currently available and is about to transmit data for the $j$\ts{th} traditional device at the $t$\ts{th} time slot, i.e., $\sum_{j=1}^Jb_{j,m}(t)=0$ and $a_{m}(t)=I+j$, the $m$\ts{th} channel will be released after $T_j-1$ time slots counting from the $(t+1)$\ts{th} time slot, i.e., $b_{j,m}(t+1)=T_j-1$; finally, if the $m$\ts{th} channel is available and not going to transmit data for the $j$\ts{th} traditional device, i.e., $\sum_{j=1}^Jb_{j,m}(t)=0$ and $a_{m}(t)\neq I+j$, $b_{j,m}(t+1)$ remains 0. To summary, we have
\begin{align}\label{trans:b}
	\!\!\!b_{j,m}(t\!+\!1)\!\!=\!\!\left\{\begin{array}{ll}
		\!\!\!\!b_{j,m}(t)\!-\!\!1&\!\!\!\!b_{j,m}\!(t)\!>\!0, \!\\
		\!\!\!\!0&\!\!\!\!b_{j,m}(t)\!=\!0,\sum_{j'\neq j}b_{j',m}(t)\!>\!0\!\\
		\!\!\!\!T_j-1&\!\!\!\!\sum_{j=1}^J\!b_{j,m}(t)\!=\!0,a_m(t)\!=\!I\!+\!j\!\\
	    \!\!\!\!0&\!\!\!\!\sum_{j=1}^J\!b_{j,m}(t)\!=\!0,a_m(t)\!\neq\!I\!+\!j.\!
	\end{array}\right.
\end{align}

\subsubsection{Reward}The reward of the whole system consists of the AoIIs at $I$ monitoring devices and the throughputs of $J$ traditional devices. Specifically, the throughput of transmitting data for the $j$\ts{th} traditional device over the $m$\ts{th} channel at the $t$\ts{th} time slot is computed as 
\begin{align}
	u_{j,m}(t)\triangleq W_m\log\left(1+\frac{g_{j,m}(t)P}{N}\right),
\end{align}
where $W_m$ is the bandwidth of the $m$\ts{th} channel, $P$ is the transmission power at the transmitters of the traditional devices, and $N$ is the noise power at the receiver of the BS. Notably, the above throughput exists only if $\mathbbm{1}_{I+j}\left(a_m(t)\right)\!\!=\!\!1$ or $b_{j,m}(t)\!\!>\!\!0$ holds, when the $m$\ts{th} channel starts to transmit data or has been reserved for transmitting data for the $j$\ts{th} traditional device at the $t$\ts{th} time slot. The reward of the whole system is defined as the summation of all the AoIIs and throughputs at the $t$ time slot, i.e.,
\begin{align}
    \begin{split}\label{reward_p1}
    	r(t)\triangleq&-\sum_{i=1}^Iw_ix_i(t)+\sum_{j=1}^Jw_{I+j}\\
        &\sum_{m=1}^M\left(\mathbbm{1}_{I+j}\left(a_m(t)\right)+\mathcal{I}\left(b_{j,m}(t)\right)\right)u_{j,m}(t),
    \end{split}   
\end{align}
where $w_i$, $i\in\{1,2,\cdots,I+J\}$, measures the importance of the $i$\ts{th} device, and the indicator function $\mathcal{I}(x)$ equals 1 if $x$ is positive and otherwise, it equals 0. 

\subsection{Problem formulation}

{\flushleft{\it \ \ 1) Markov decision problem formulation}}

This paper aims to maximize the long-term average reward for \eqref{reward_p1}. However, the objective $\max_{\{\bm{a}(t)\}}\lim_{T\rightarrow\infty}$ $\mathbb{E}_{\text{Pr}\{\bm{x}'|\bm{x},\bm{a}\},\text{Pr}\{\bm{G}'|\bm{G}\}}\left[\frac{1}{T}\sum_{t=1}^Tr(t)\right]$ may not exist under some assignments of $\{\bm{a}(t)\}$ (cf. \cite[Example 8.1.1]{puterman}), where the expectation is taken with respect to the {\it AoII vector} and the {\it channel gains}. Therefore, we utilize {\it liminf} average optimality criteria and formulate the joint scheduling problem between the task-oriented and data-oriented communications as 
\begin{align*}
	\textbf{(P1)}\max_{\{\bm{a}(t)\}}\ \ &\liminf_{T\rightarrow\infty}\mathbb{E}_{\text{Pr}\{\bm{x}'|\bm{x},\bm{a}\},\text{Pr}\{\bm{G}'|\bm{G}\}}\left[\frac{1}{T}\sum_{t=1}^Tr(t)\right]\\
	\text{s.t.}\ \ &\eqref{con:P},\eqref{trans:x1},\eqref{trans:x2},\eqref{trans:x3},\eqref{trans:g},\eqref{trans:b}.
\end{align*}
To solve problem {\bf (P1)}, we need to find the optimal policy $\boldsymbol{\pi}\triangleq\left(\pi_1,\pi_2,\cdots,\pi_t,\cdots\right)$, where $\pi_t$ is the optimal decision rule at the $t$\ts{th} time slot and maps the history of the states and actions $h(t)\triangleq(\bm{s}(1),\bm{a}(1),\cdots,\bm{s}(t-1),$ $\bm{a}(t-1),\bm{s}(t))$ to the optimal distribution of the current action $\bm{a}(t)$, i.e., $\pi_t: \mathcal{H}(t)\times\mathcal{A}_{\bm{s}(t)}\rightarrow[0,1]$ with $\mathcal{H}(t)$ being the set of all histories $h(t)$.
 
It can be checked that problem {\bf (P1)} has upper-bounded rewards and countable states, and thus the stationary policies achieving its optimum exist if certain conditions are satisfied \cite{det_sta}. Then, we obtain the following proposition.
\begin{Prop}\label{prop1}
    There exist stationary policies to achieve the optimal value of problem {\bf (P1)}.
\end{Prop}
\begin{IEEEproof}
Please see Appendix \ref{proof1}.
\end{IEEEproof}

Based on Proposition \ref{prop1}, the {\it liminf} average optimality criteria for problem {\bf (P1)} can be replaced by {\it lim}, since the Cesaro limit always exists for stationary policies \cite{puterman}. Moreover, the state occurrence probabilities are constants under stationary policies \cite{puterman}, where the state occurrence probability of state $\bm{s}$ is defined as $\lim_{T\rightarrow\infty}\sum_{t=1}^{T}\mathbbm{1}_{\bm{s}}(\bm{s}(t))/T$ \cite{puterman}. Based on this property, we can simplify problem {\bf (P1)} to an equivalent Markov decision problem {\bf (P2)}, which is given as
\begin{itemize}
	\item {\it State:} $\hat{\bm{s}}(t)\triangleq(\bm{x}(t),\bm{G}(t),\bm{b}(t))$, where $\bm{b}(t)\triangleq[b_1(t),$ $b_2(t),\cdots,b_M(t)]^T\in\mathbb{Z}_{\geq0}^{M\times1}$ is defined as the {\it channel release time}. Here, $b_m(t)$ is equal to the number of the remaining time slots for the release of the $m$\ts{th} channel and defined as $b_m(t)\triangleq\sum_{j=1}^Jb_{j,m}(t)$, $m\in\{1,2,\cdots,M\}$. Apparently, $\bm{b}(t)$ can be derived from $\bm{B}(t)$ and thus $\hat{\bm{s}}(t)$ can be derived from $\bm{s}(t)$. The new state space is $\hat{\mathcal{S}}\triangleq\mathbb{Z}_{\geq0}^{I\times1}\times\mathcal{G}^{J\times M}\times\mathcal{b}$, where $\mathcal{b}$ is the value space of $\bm{b}(t)$;
	\item {\it Action:} $\bm{a}(t)$, which is constrained by \eqref{con:P};
	\item {\it Transitions:} \eqref{trans:x1}, \eqref{trans:x2}, \eqref{trans:x3}, \eqref{trans:g}, and
	\begin{align}\label{trans:b_new}
	    \!\!\!b_m(t+1)\!=\!\left\{\begin{array}{ll}
		    \!\!b_m(t)\!-\!1&\!b_m(t)\!>\!0\\
		    \!\!T_{a_m(t)-I}\!-\!1&\!b_m(t)\!=\!0,\!a_m(t)\!>\!I\\
		    \!\!0&\!b_m(t)\!=\!0,\!a_m(t)\!\leq\!I;
	    \end{array}\right.	    	
	\end{align}
	where \eqref{trans:b_new} is derived by \eqref{trans:b} and $b_m(t)\triangleq\sum_{j=1}^Jb_{j,m}(t)$.
    \item {\it Reward:} $\hat{r}(t)$ is defined as
    \begin{align}\label{def:hatr}
    	\hat{r}(t)\!\triangleq\!-\!\sum_{i=1}^Iw_ix_i(t)\!+\!\!\sum_{j=1}^J\!w_{I+j}\!\!\sum_{m=1}^M\!\!\mathbbm{1}_{I+j}\left(a_m(t)\right)\bar{u}_{j,m}(t),   	
    \end{align} where, $\bar{u}_{j,m}(t)$ is defined as {\normalfont
    \begin{align}\label{def:baru}
    	\bar{u}_{j,m}(t)\triangleq\sum_{\tau=t}^{t+T_j-1}\mathbb{E}_{\text{Pr}\{\bm{G}'|\bm{G}\}}\left[u_{j,m}(\tau)|_{g_{j,m}(t)}\right];
    \end{align}}
    \item {\it Problem formulation:} with the {\it lim} average optimality criteria, problem {\bf (P2)} is reformulated as
    \begin{align}
        &\textbf{(P2)}\max_{\pi\in\Pi^{S}}\lim_{T\rightarrow\infty}\!\!\!\mathbb{E}_{\pi,\text{Pr}\{\bm{x}'|\bm{x},\bm{a}\},\text{Pr}\{\bm{G}'|\bm{G}\}}\left[\frac{1}{T}\sum_{t=1}^T\hat{r}(t)\right]\label{P2}\\
        &\qquad\quad\text{s.t.}\ \ \eqref{con:P},\eqref{trans:x1},\eqref{trans:x2},\eqref{trans:x3},\eqref{trans:g},\eqref{trans:b_new},\nonumber
    \end{align}where $\Pi^{S}$ is the set of all stationary policies and the expectation is taken with respect to the policy $\pi$, the {\it AoII vector}, and the {\it channel gains}.
\end{itemize}

Apparently, problem {\bf (P2)} has a much simpler structure than problem {\bf (P1)}, where the dimension of the state space is reduced from $I+2JM$ to $I+JM+M$, the transitions in \eqref{trans:b_new} evolves much simper than the ones in \eqref{trans:b}, and the reward in $\hat{r}(t)$ no longer involves the high-dimension matrix $\bm{B}(t)$ compared with the reward in $r(t)$. Moreover, compare problems {\bf (P1)} and {\bf (P2)}, we obtain the following proposition.
\begin{Prop}\label{prop2}
Problem {\bf (P2)} is equivalent to problem {\bf (P1)}. Particularly, for any stationary policy optimizing problem {\bf (P1)}, there exists another stationary policy that optimizes problem {\bf (P2)}, and the inverse holds, too. Moreover, the optimal values of the two problems are the same.
\end{Prop}
\begin{IEEEproof}
Please see Appendix \ref{proof2}.
\end{IEEEproof}
However, problem {\bf (P2)} is still difficult to be solved since it is a Markov decision problem with the {\it lim} average optimality criteria, and the existing algorithms addressing such problems, e.g., relative value iteration algorithm \cite{dp1}, still suffer from the curse of dimensionality. To tackle this issue, we refer \cite[Proposition 4.1.7]{dp2} and verify the existence of the Blackwell policies in problem {\bf (P2)}. If the Blackwell policies do exist, the optimal policies of problem {\bf (P2)} can be found by solving a discounted version of it, which is much easier to be addressed. Particularly, the discounted version of problem {\bf (P2)} is given as
\begin{align}
    &\textbf{(P3)}\max_{\pi\in\Pi^{S}}\ \lim_{T\rightarrow\infty}\mathbb{E}_{\pi,\text{Pr}\{\bm{x}'|\bm{x},\bm{a}\},\text{Pr}\{\bm{G}'|\bm{G}\}}\left[\sum_{t=1}^T\alpha^{t-1}\hat{r}(t)\right]\label{P2}\\
    &\qquad\quad\text{s.t.}\quad\eqref{con:P},\eqref{trans:x1},\eqref{trans:x2},\eqref{trans:x3},\eqref{trans:g},\eqref{trans:b_new},\nonumber
\end{align}
where $\alpha\in(0,1)$ is the discount factor. Then, we derive the relationship between problems {\bf (P2)} and {\bf (P3)} in the following proposition.
\begin{Prop}\label{prop3}
When the discount factor $\alpha\in(0,1)$ is sufficiently close to 1, there exist stationary policies to simultaneously achieve the optimal values of problems {\bf (P2)} and {\bf (P3)}.
\end{Prop} 
\begin{IEEEproof}
Please see Appendix \ref{proof3}.
\end{IEEEproof}
\begin{Rem}
With Proposition \ref{prop3}, we can solve problem {\bf (P2)} by first selecting a proper discount factor $\alpha$ and then deriving a stationary optimal policy of problem {\bf (P3)} with the chosen discount factor. However, how to solve problem {\bf (P3)} is still challenging due to its three features: 1) large state space; 2) large and discrete action space; and 3) multiple time-varying action constraints in \eqref{con:P}. However, conventional approaches including dynamic programming {\normalfont\cite{dp1}} and Lyapunov drift optimization {\normalfont\cite{neely}}, can barely deal with Markov decision problems with the first two features. Modern DRL algorithms {\normalfont\cite{sutton}} can neither efficiently address the Markov decision problems with large and discrete action space{\footnote{The most advanced algorithm to address Markov decision problem with large discrete action space is the Wolpertinger policy \cite{wolper}, which adds an action-embedding module right after the deep deterministic policy gradient (DDPG\textsuperscript{\cite{ddpg}}) algorithm and directly discretizes the continuous-valued action generated by DDPG. However, the Wolpertinger policy has a poor interpretability and could generate really large training variance even when the action space is small \cite{wolper}.}}, and most of them deploy the trial and error mechanism {\normalfont\cite{sutton}}, where the trial part would always violate the time-varying constraints in \eqref{con:P} and thus the training procedure would be terminated.
\end{Rem}

{\flushleft{\it \ \ 2) Markov game formulation}}

To overcome the three challenges in problem {\bf (P3)}, we treat each channel as one virtual agent, which can first observe the system state and then independently determine the device for data transmission over itself. Therefore, problem {\bf (P3)} can be reformulated as an equivalent Markov game {\bf (P4)}, which has the same state, transition, and optimality criteria as those of problem {\bf (P3)} and also contains
\begin{itemize}
	\item {\it Observation $\bm{s}_m(t)\triangleq(\bm{x}(t),\bm{g}_m(t),b_m(t))$ at the $m$\ts{th} agent:} $\bm{g}_m(t)$ contains the channel gains for transmitting data for $J$ traditional devices over the $m$\ts{th} channel, respectively, and is defined by $\bm{g}_m(t)\triangleq[g_{1,m}(t),$ $g_{2,m}(t),\cdots,g_{J,m}(t)]^T$;
	\item {\it Action $a_m(t)$ at the $m$\ts{th} agent:} it is now individually constrained by $b_m(t)a_m(t)=0$;
	\item {\it Reward $r_m(t)$ for the $m$\ts{th} agent:} it is set to be $\hat{r}(t)$ in \eqref{def:hatr}.
\end{itemize}
\begin{Rem}\label{rem21}
On the design of the agent observation, although the whole state $\hat{\bm{s}}(t)$ is observable for each agent, only $\bm{s}_m(t)$ is reserved as the $m$\ts{th} agent's observation. The reasons are: 1) the dimensionality of $\hat{\bm{s}}(t)$ is too large and difficult to be dealt with; 2) except the elements in $\bm{s}_m(t)$, the rest ones in $\hat{\bm{s}}(t)$ are weekly correlated to the $m$\ts{th} agent; 3) the union of $\bm{s}_m(t)$ among all agents covers all the elements in state $\hat{\bm{s}}(t)$ and thus the algorithms developed based on $\bm{s}_m(t)$ can achieve the same optimum with the algorithms based on $\hat{\bm{s}}(t)$ (see {\normalfont Fig.~\ref{plot_1_2}} in {\normalfont Section \ref{Section4}}).
\end{Rem}

Markov game {\bf (P4)} is obviously equivalent to problem {\bf (P3)} since all its agents cooperatively optimize the common reward $\hat{r}(t)$ in problem {\bf (P3)}. Meanwhile, it faces similar challenges to problem {\bf (P3)}. Fortunately, the existing multi-agent reinforcement learning (MARL) algorithms can well address the challenge from the large state space issue by approximating the optimal policy with neural network (NN), and relieve the challenge from the large discrete action space issue by deploying decentralized learning mechanism, where each agent needs only to determine its own action in a much smaller action space $\{0,\!1,\cdots,I+J\}$. 

However, there are still two obstacles for MARL algorithms to solve {\bf (P4)}: 1) the action space for each agent, i.e., $\{0,1,\cdots,I+J\}$, is still large and would slow down the training procedure for MARL algorithms to a great extent (see Fig.~\ref{plot_1_2} in Section \ref{Section4}); 2) the action space for each agent is constrained by one time-varying constraint, which again collides with the trial and error mechanism deployed in MARL algorithms; 

\section{Whittle's Index Guided Multi-Agent Proximal Policy Optimization}\label{Section3}
To address Markov game {\bf (P4)}, we propose WI-MAPPO, which mainly comprises one Whittle's index guided action fusion (WIAC) module and multiple proximal policy optimization (PPO) modules. Particularly, the WIAC module calculates the Whittle's index for $I$ monitoring devices based on their AoIIs, and accordingly determines the priorities for their data transmissions. Then, all agents only need to transmit data for the group of monitoring devices with the highest priority, and thus the action space for each individual agent is greatly shrunk. Meanwhile, we modify both the actor network and the probability ratio derivation of the training algorithm for multi-agent proximal policy optimization (MAPPO) to train the proposed WI-MAPPO, which perfectly addresses the time-varying constraint issue.

In the following, we first introduce the structure of the proposed algorithm. Then, we present the offline training algorithm in details. Finally, we briefly introduce the online applying algorithms for solving Markov game {\bf (P4)}.

\subsection{Structure of proposed algorithm}
\begin{figure*}[!htb]
\centering
\includegraphics[width=6.8in]{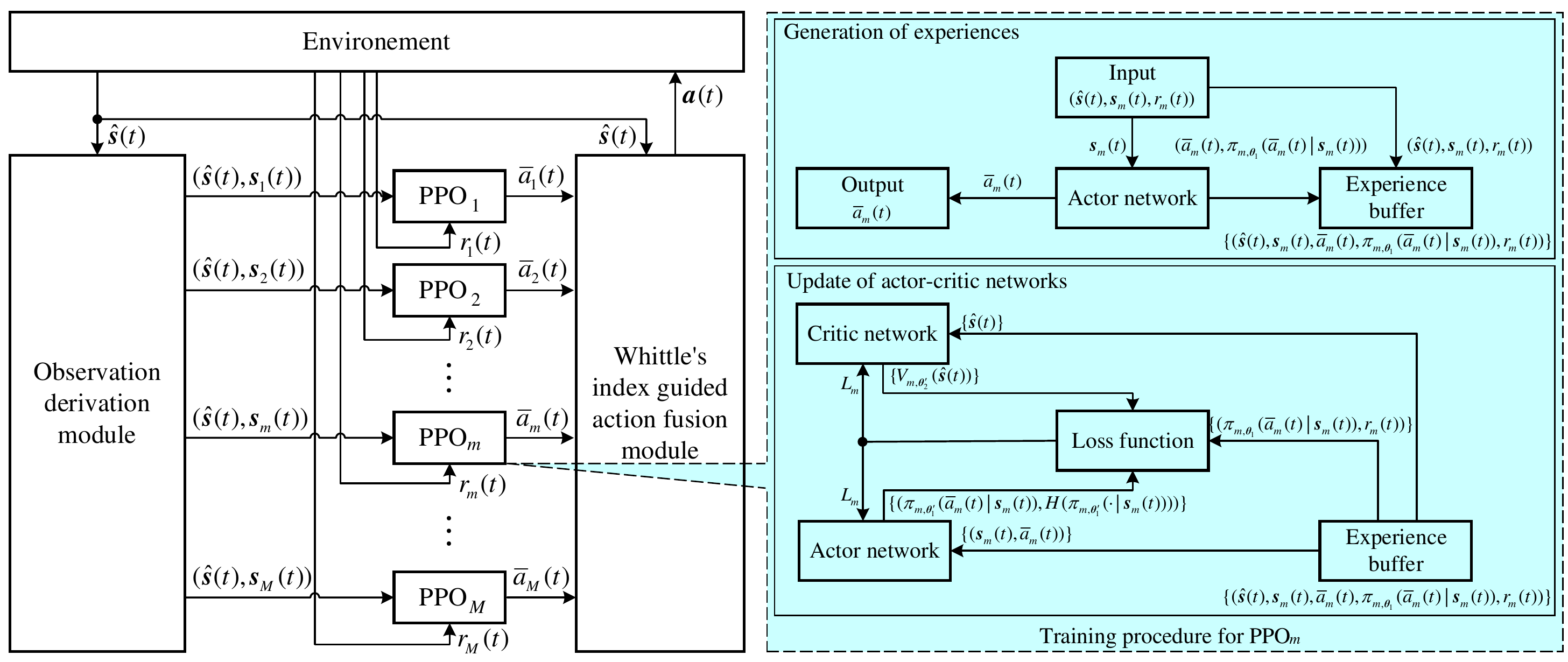}
\caption{Structure of the proposed WI-MAPPO algorithm.}\label{fig7}
\end{figure*}

As illustrated in Fig.~\ref{fig7}, the main body of the proposed algorithm consists of one observation derivation (OD) module, $M$ PPOs, and one WIAC module.

\subsubsection{OD module}
This module derives the agent observations $\bm{s}_1(t)$, $\bm{s}_2(t)$, $\cdots$, $\bm{s}_M(t)$ from the state $\hat{\bm{s}}(t)$ based on the definition $\bm{s}_m(t)\triangleq(\bm{x}(t),\bm{g}_m(t),b_m(t))$.

\subsubsection{PPOs}
Each agent utilizes a PPO module to determine the device for data transmission over its channel. Specifically, The PPO utilized by the $m$\ts{th} agent is named as PPO$_m$ and it has a simple structure:
\begin{itemize}
	\item Actor network: It contains a fully connected NN parameterized by $\boldsymbol{\theta}_1$. Particularly, this NN takes $\bm{s}_m(t)$ as the input and thus has $I+J+1$ nodes at the input layer; the output layer has $J+2$ nodes and represents the probabilities of transmitting data for all $J$ traditional devices, transmitting data for one monitoring device, and not starting new transmission, respectively, which are denoted by $\pi_{m,\boldsymbol{\theta}_1}(1|\bm{s}_m(t))$, $\pi_{m,\boldsymbol{\theta}_1}(2|\bm{s}_m(t))$, $\cdots$, and $\pi_{m,\boldsymbol{\theta}_1}(J+2|\bm{s}_m(t))$. The output of the actor network, which is named as the PPO action, is denoted as $\bar{a}_m(t)\in\{1,2,\cdots,J+2\}$. Particularly, when the $m$\ts{th} channel is currently available, i.e., $b_m(t)=0$, $\bar{a}_m(t)$ is equal to a discrete random variable $X$, whose probability mass function is $\text{Pr}\{X=j\}=\pi_{m,\boldsymbol{\theta}_1}(j|\bm{s}_m(t)),\ \forall j\in\{1,2,\cdots,J+2\}$. And when the $m$\ts{th} channel is occupied, i.e., $b_m(t)>0$, $\bar{a}_m(t)$ is equal to $J+2$. To summary, we have
	\begin{align}\label{def:bara}
		\bar{a}_m(t)=\left\{\begin{array}{ll}
			J+2&b_m(t)>0\\			
			X&b_m(t)=0.
		\end{array}\right.
	\end{align}
	Remarkably, trivial action selection in the actor network of PPO simply follows the second line in \eqref{def:bara} \cite{ppo}. While in our design, we introduce the first line of \eqref{def:bara} to manually change the action $\bar{a}_m(t)$ to not starting new transmission, i.e., $\bar{a}_m(t)=J+2$, when the $m$\ts{th} channel is currently occupied. Such modification will benefit the offline training procedure discussed in \ref{offline_section} and we will explain the reasons in Remark \ref{rem33}.
	\item Critic network: It contains a fully connected NN parameterized by $\boldsymbol{\theta}_2$. Specifically, this NN takes $\hat{\bm{s}}(t)$ as the input and thus has $I+JM+M$ nodes at the input layer; the output layer of the NN has only one node and gives the estimation on the maximum total discounted reward starting from state $\hat{\bm{s}}(t)$, which is also called as the value function of state $\hat{\bm{s}}(t)$ and denoted by $V_{m,\boldsymbol{\theta}_2}(\hat{\bm{s}}(t))$,;
	\item Experience buffer: It stores the experiences generated in the offline training procedure, where the experiences are the five-component tuples $(\hat{\bm{s}}(t),\bm{s}_m(t),\bar{a}_m(t),\pi_{m,\boldsymbol{\theta}_1}(\bar{a}_m(t)|\bm{s}_m(t)),r_m(t))$.
\end{itemize}

\begin{Rem}\label{rem31}
On the design of the actor network, it is reasonable to set the output dimension of the actor network's NN as $I+J+1$ and let the output values represent the probabilities of transmitting data for all $I+J$ devices, and not starting new transmission, respectively, at state $\bm{s}_m(t)$. By doing so, the $m$\ts{th} agent can directly determine its action by sampling from the action distribution generated by the trained actor network's NN. However, high dimensionality of the output, which is equal to $I+J+1$, requires large NN, and the convergence of the training procedure among multiple agents might be very difficult. In the proposed algorithm, we design the WIAC module to figure out the group of monitoring device with the highest potential to minimize their long-term average AoIIs, by which each agent only needs to determine whether it is willing to transmit data for one monitoring device. If it is, the agent will select any one in the group figured out by the WIAC module for data transmission. Therefore, the actor network in the proposed algorithm only requires an output with $J+2$ dimensions.
\end{Rem}

\subsubsection{WIAC module}\label{wiac}
This module determines the action $\bm{a}(t)$ based on state $\hat{\bm{s}}(t)$ and $M$ PPO actions $\bar{\bm{a}}(t)\triangleq[\bar{a}_1(t),\bar{a}_2(t),\cdots,\bar{a}_M(t)]^T$. In this subsection, we first introduce the design intuition of this module, and then introduce the explicit method to compute $\bm{a}(t)$. 

\textbf{Design intuition:} WIAC module first solves the following problem {\bf (P5)}: {\it Given the current AoII values at $I$ monitoring devices, which group of the monitoring devices should be selected for status update transmissions over limited number of channels to minimize the long-term average weighted AoIIs?}\\
Problem {\bf (P5)} is a restless multi-armed bandit (RMAB) problem and a typical Whittle's index algorithm is deployed to solve it \cite{whittle}. Particularly, we first model $I$ monitoring devices as $I$ individual agents and the goal of each agent is to minimize its own long-term average AoII. Then, by studying the AoII evolution of each agent, the maximum offer that each agent is willing to pay for hiring one channel at the current time slot can be derived, which is named as the Whittle's index for this agent. Finally, the group of agents with the highest Whittle's indices will be selected for data transmissions. Remarkably, the Whittle's index algorithm is validated to be a near-optimal algorithm to solve RMABs \cite{whittle}. 

However, to apply this algorithm, the existence of the Whittle's index should be guaranteed. Therefore, in the following, we first validate the existence of the Whittle's index for problem {\bf (P5)}, and then derive a Whittle's index table with an exhausted search algorithm. Finally, to derive the final action $\bm{a}(t)$, we first check the values of $\bar{\bm{a}}(t)$ and obtain the number of channels willing to transmit data for one monitoring device, i.e., $\sum_{m=1}^M\mathbbm{1}_{J+1}(\bar{a}_m(t))$. Then, we look up the Whittle's index table and transmit data for the group of the monitoring devices with the highest Whittle's indices over these channels, where the size of this group is also equal to $\sum_{m=1}^M\mathbbm{1}_{J+1}(\bar{a}_m(t))$.

\textbf{Existence of Whittle's index:} The following proposition validates the existence of the Whittle's index for problem {\bf (P5)}.
\begin{Prop}\label{prop4}
There exists Whittle's index for problem {\bf (P5)}. 
\end{Prop}
\begin{IEEEproof}[Sketch of proof]
To validate the existence of Whittle's index, we first decouple problem {\bf (P5)} to $I$ sub-problems, where the $i$\ts{th} sub-problem is to minimize the average AoII at the $i$\ts{th} monitoring device. Next, we analyze the properties of these sub-problems and validate that the optimal policies for these sub-problems are of threshold type. Then, based on the ``threshold" feature on the optimal policy, we prove the indexability for the decoupled sub-problems, which validates the existence of the Whittle's index for problem {\bf (P5)}. Please check Appendix~\ref{proof4} for more details.
\end{IEEEproof}

\textbf{Derivation of Whittle's index table:}
The Whittle's index for the $i$\ts{th} monitoring device with its AoII being $x$, notated as $I_i(x)$, is defined as the additional cost $C$ that makes both transmitting data and not transmitting data for the $i$\ts{th} monitoring device equally desirable, i.e., 
\begin{align}\label{def:whittle}
f_i(x,C)=f_i(x+1,C),	
\end{align}
where the additional cost $C$ and the average cost function $f_i(x,C)$ are introduced in Appendix \ref{proof4}. Since to derive the closed-form formulation of $I_i(x)$ by solving \eqref{def:whittle} is very difficult, we use exhausted search algorithm to obtain $I_i(x)$ with the searching step and searching area as $\Delta c$ and $[C_L,C_U]$, respectively. Moreover, by exploiting the fact that $I_i(x)$ is non-decreasing with respect to $x$ \cite{whittle}, the above exhausted searching algorithm can be improved. Notably, we would generate a sufficiently large Whittle's index table by executing this algorithm before the offline training procedure.

\textbf{Derivation of $\bm{a}(t)$:} During the offline training procedure, WIAC module first counts the number of agents willing to transmit data for monitoring devices, i.e., the agents satisfying $\bar{a}_m(t)=J+1,\ m\in\{1,2,\cdots,M\}$. Particularly, this number at the $t$\ts{th} time slot is denoted as $A(t)$ and defined as $A(t)\triangleq\sum_{m=1}^M\mathbbm{1}_{J+1}(\bar{a}_m(t))$. Next, WIAC module looks up the generated Whittle's index table and obtains the Whittle's indices for $I$ monitoring devices according to their current AoII values $\bm{x}(t)$. Then, the index of the monitoring device with the $l$\ts{th} highest Whittle's index is denoted as $W_l(t)\in\{1,2,\cdots,I\}$ and $A(t)$ monitoring devices with the top $A(t)$ Whittle's indices are picked out. Finally, the $A(t)$ agents satisfying $\bar{a}_m(t)=J+1,\ m\in\{1,2,\cdots,M\}$ transmit data for the picked out $A(t)$ monitoring devices over their channels and accordingly the action $\bm{a}(t)=(a_1(t),a_2(t),\cdots,a_M(t))$ is computed as
\begin{align}\label{def:hata}
    a_m(t)=\left\{\begin{array}{ll}
    	I+\bar{a}_m(t)&\bar{a}_m(t)<J+1\\
    	W_{l_m}(t)&\bar{a}_m(t)=J+1\\
    	0&\bar{a}_m(t)=J+2,
    \end{array}\right.
\end{align}
where $l_m$ is the number of elements equaling $J+1$ in $[\bar{a}_1(t),\bar{a}_2(t),\cdots,\bar{a}_m(t)]^T$.

\subsection{Offline training}\label{offline_section} 
Based on the historical observed samples, we can easily approximate the values of $\text{Pr}\{\bm{G}'|\bm{G}\}$, $p_i$, and $q_i$, and then simulate an offline environment accordingly. Finally, we develop the offline training algorithm by interacting with it.

\subsubsection{Offline environment simulation}\label{off_env}
To mimic the real environment, the offline environment needs to fulfill two functions:
\begin{itemize}
	\item State evolution: Given $\hat{\bm{s}}(t)=(\bm{x}(t),\bm{G}(t),\bm{b}(t))$ and $\bm{a}(t)$, we first simulate $\bm{x}(t+1)$ based on \eqref{trans:x1}, \eqref{trans:x2}, and \eqref{trans:x3}, and simulate $\bm{G}(t+1)$ based on \eqref{trans:g} and the approximated $\text{Pr}\{\bm{G}'|\bm{G}\}$, $p_i$, and $q_i$. Then, we directly compute $\bm{b}(t+1)$ based on \eqref{trans:b_new}. Thus, $\hat{\bm{s}}(t+1)$ is obtained;
	\item Reward generation: Given $\hat{\bm{s}}(t)$ and $\bm{a}(t)$, we compute $r_1(t)$, $r_2(t)$, $\cdots$, $r_M(t)$ based on \eqref{def:hatr}, \eqref{def:baru} and the fact $r_m(t)=\hat{r}(t)$.
\end{itemize}
\subsubsection{Offline training}
As illustrated in the right part of Fig.~\ref{fig7}, we alternatingly generate experiences by deploying the latest actor-critic network and update the actor-critic network according to the latest generated experiences. We specific these two steps as follows.
\begin{itemize}
	\item Generation of experiences: First, obtain the $\bm{s}_1(t)$, $\cdots$, $\bm{s}_M(t)$ from the observed $\hat{\bm{s}}(t)$ based on OD module. Next, by utilizing the actor networks for PPOs, obtain PPO action $\bar{\bm{a}}(t)$ and the value of $\pi_{m,\boldsymbol{\theta}_1}(\bar{a}_m(t)|\bm{s}_m(t))$ for all $m\in\{1,2,\dots,M\}$ according to \eqref{def:bara}. Then, with the known $\bar{\bm{a}}(t)$ and $\hat{\bm{s}}(t)$, derive action $\bm{a}(t)$ by using the WIAC module. Finally, obtain the agent rewards $r_1(t),\cdots,r_M(t)$ by interacting with the offline environment. We pack the above information as experiences $e_1(t), e_2(t), \cdots, e_M(t)$, where $e_m(t)$ is defined as
	\begin{align}\label{pack}
		e_m(t)\!\triangleq\!(\hat{\bm{s}}(t),\!\bm{s}_m(t),\!\bar{a}_m(t),\pi_{m,\boldsymbol{\theta}_1}(\bar{a}_m(t)|\bm{s}_m(t)),r_m(t)),
	\end{align}
    and then we store $e_m(t)$ in the experience buffer in PPO$_m$. Remarkably, we can continuously generate $N_B$ experiences for each PPO before the update of the actor-critic networks and empty all the experience buffers after each update.
    \item Updation of actor-critic networks: Each updation performs $N_U$ epochs of optimization on the generated $N_B$ experiences and each epoch would modify the parameters of the actor-critic networks for all PPOs. Particularly, at the beginning of each epoch, we first denote the actor and critic networks for PPO$_m$ at this moment as $\pi_{m,\boldsymbol{\theta}_1'}$ and $V_{m,\boldsymbol{\theta}_2'}$, respectively, Next, we estimate $N_B$ value functions as
    \begin{align}\label{def:vm}
        V_m\!(t,\!\hat{\bm{s}}(t))\!\!=\!\!r_m\!(t)\!+\!\alpha r_m\!(t+1)\!+\!\cdots\!+\!\alpha^{N_B}\!r_m\!(N_B\!)
    \end{align}
        with $t\in\{1,2,\cdots,N_B\}$, where $\{V_m(t,\hat{\bm{s}}(t))\}_{t=1}^{N_B}$ are the value functions of states $\{\hat{\bm{s}}(t)\}_{t=1}^{N_B}$ for PPO$_m$ \cite{ppo}. Then, we utilize the current critic network $V_{m,\boldsymbol{\theta}_2'}$ to estimate $N_B$ advantage functions as
    \begin{align}\label{def:am}
        A_m(t)\!\!=\!\!V_m(t,\!\hat{\bm{s}}(t))\!-\!V_{m,\boldsymbol{\theta}_2'}\!(\hat{\bm{s}}(t)),\!t\!\in\!\!\{1,2,\!\cdots\!,\!N_B\!\},
    \end{align}
        where $\{A_m(t)\}_{t=1}^{N_B}$ are the advantage functions for PPO$_m$ \cite{ppo}, and utilize the current actor network $\pi_{m,\boldsymbol{\theta}_1'}$ to derive $N_B$ probability ratios as
    \begin{align}\label{def:rm}
        R_m(t)=\left\{\begin{array}{ll}
        	1&b_m(t)>0\\
        	\frac{\pi_{m,\boldsymbol{\theta}_1'}(\bar{a}_m(t)|\bm{s}_m(t))}{\pi_{m,\boldsymbol{\theta}_1}(\bar{a}_m(t)|\bm{s}_m(t))}&b_m(t)=0,
        \end{array}\right.
    \end{align}
    for all $t\in\{1,2,\cdots,N_B\}$. Finally, the surrogate loss for PPO$_m$ is computed as \cite{ppo}
    \begin{align}
        \begin{split}\label{def:surro}
        	L_m=&\sum_{t=1}^{N_B}\frac{1}{N_B}\Big(\!-\text{min}\left(R_m(t)A_m(t),\right.\\
        	&\left.\left.\text{clip}(R_m(t),1-\epsilon,1+\epsilon)A_m(t)\right)\right.\\
        	&+c_1(V_m(t,\hat{\bm{s}}(t))-V_{m,\boldsymbol{\theta}_2'}(\hat{\bm{s}}(t)))^2\\
        	&-c_2H\left(\pi_{m,\boldsymbol{\theta}_1'}(\cdot|\bm{s}_m(t))\right)\!\Big),
        \end{split}       
    \end{align}
    where $\text{clip}(x,a,b)\triangleq\text{min}(\text{max}(x,a),b)$ clamps $x$ into the area $[a,b]$; $H(\pi_{m,\boldsymbol{\theta}_1'}(\cdot|\bm{s}_m(t))$ is the entropy of the stochastic output generated by the current actor network $\pi_{m,\boldsymbol{\theta}_1'}$ with $\bm{s}_m(t)$ as input; $\epsilon$, $c_1$, and $c_2$ are some constants. Particularly, the first term in \eqref{def:surro} is a pessimistic bound, which could improve the actor-critic networks for PPO in a considerably stable manner; the second MSE term is essential for the convergence of the training of the actor-critic network \cite{ppo}; and the last term adopts an entropy bonus to ensure sufficient exploration. Remarkably, both the actor and critic networks backpropagate this surrogate loss to update their parameters $\boldsymbol{\theta}_1'$ and $\boldsymbol{\theta}_2'$.
\end{itemize}
The details of the offline training algorithm are summarized in Algorithm~\ref{offline}.
\begin{algorithm}[ht]
\caption{Offline training algorithm for joint scheduling}\label{offline}
\begin{algorithmic}[1]
\STATE Randomly initialize the actor-critic networks PPO$_1$, PPO$_2$, $\cdots$, and PPO$_M$;
\STATE Initialize one experience buffer for each PPO;
\STATE Input the values of $\text{Pr}\{\bm{G}'|\bm{G}\}$, $\{p_i\}_{i=1}^I$, $\{q_i\}_{i=1}^I$, $\{w_i\}_{i=1}^{I+J}$, $\{W_m\}_{m=1}^M$, $P/N$, $\alpha$, $\Delta c$, $C_L$, $C_U$, $N_B$, $N_U$, $\epsilon$, $c_1$, $c_2$;
\STATE Derive the Whittle's index table by executing the exhausted search algorithm specified in \ref{wiac};
\STATE Generate the offline environment based on \ref{off_env}; 
\STATE \textbf{for} $\text{episode}=1,2,\cdots$
\STATE \ \ \ Let $\bm{x}(1)=0^{I\times 1}$ and $\bm{b}(1)=0^{M\times 1}$. Let $\bm{G}(1)$ be any
\STATEx\ \ \ element in $\mathcal{G}^{J\times M}$;
\STATE \ \ \ $\hat{\bm{s}}(1)=(\bm{x}(1),\bm{G}(1),\bm{b}(1))$;
\STATE \ \ \ \textbf{for} $t=1,2,\cdots,N_B$
\STATE \ \ \ \ \ \ Send $\hat{\bm{s}}(t)$ to OD module and derive $\{\bm{s}_m(t)\}_{m=1}^M$;
\STATE \ \ \ \ \ \ \textbf{for} $m=1,2,\cdots,M$
\STATE \ \ \ \ \ \ \ \ \ Send $\bm{s}_m(t)$ to the actor Network for PPO$_m$ and
\STATEx\ \ \ \ \ \ \ \ \ derive $\bar{a}_m(t)$, $\pi_{m,\boldsymbol{\theta}_1}(\bar{a}_m(t)|\bm{s}_m(t))$;
\STATE \ \ \ \ \ \ \textbf{end for}
\STATE \ \ \ \ \ \ Send $\hat{\bm{s}}(t)$ and $\bar{\bm{a}}(t)$ to the WIAC module and derive
\STATEx\ \ \ \ \ \ $\bm{a}(t)$;
\STATE \ \ \ \ \ \ Send $\hat{\bm{s}}(t)$ and $\bm{a}(t)$ to the offline environment and
\STATEx\ \ \ \ \ \ derive $\hat{\bm{s}}(t+1)$, $r_1(t)$, $\cdots$, $r_M(t)$;
\STATE \ \ \ \ \ \ \textbf{for} $m=1,2,\cdots,M$
\STATE \ \ \ \ \ \ \ \ \ Pack experience $e_m(t)$ by \eqref{pack} and store it into
\STATEx\ \ \ \ \ \ \ \ \ the experience buffer in PPO$_m$;
\STATE \ \ \ \ \ \ \textbf{end for}
\STATE \ \ \ \textbf{end for}
\STATE \ \ \ \textbf{for} $\text{epoch}=1,2,\cdots,N_U$
\STATE \ \ \ \ \ \ \textbf{for} $m=1,2,\cdots,M$
\STATE \ \ \ \ \ \ \ \ \ Load the experiences $\{e_m(t)\}_{t=1}^{N_B}$ from the
\STATEx\ \ \ \ \ \ \ \ \ experience buffer in PPO$_m$;
\STATE \ \ \ \ \ \ \ \ \ Derive $\{V_m(t,\hat{\bm{s}}(t))\}_{t=1}^{N_B}$, $\{A_m(t)\}_{t=1}^{N_B}$,
\STATEx\ \ \ \ \ \ \ \ \ $\{R_m(t)\}_{t=1}^{N_B}$ based on \eqref{def:vm}, \eqref{def:am}, \eqref{def:rm};
\STATE \ \ \ \ \ \ \ \ \ Derive $L_m$ based on \eqref{def:surro};
\STATE \ \ \ \ \ \ \ \ \ Update the actor-critic networks for PPO$_m$ by
\STATEx\ \ \ \ \ \ \ \ \ backpropagating $L_m$;
\STATE \ \ \ \ \ \ \textbf{end for}
\STATE \ \ \ \textbf{end for}
\STATE \ \ \ Empty the experience buffers for PPOs;
\STATE \textbf{end for} 
\end{algorithmic}
\end{algorithm}
\begin{Rem}\label{rem33}
Compared with the conventional training algorithm for MAPPO, the proposed training algorithm modifies the actor network for each PPO in \eqref{def:bara} and the probability ratio derivation in \eqref{def:rm}. Such modifications have two advantages:
\begin{itemize}
	\item The proposed training algorithm will not violate the time-varying constraints in \eqref{con:P} nor terminate the training procedure. The conventional training algorithm for MAPPO selects action by sampling from the action distribution generated by the actor network, and thus may select infeasible ones. However, based on the modification in \eqref{def:bara}, the proposed algorithm forces the agent not to start new data transmission when its channel is occupied, by which the generated action never violates the constraints in \eqref{con:P};
	\item The proposed training algorithm perfectly extends the conventional training algorithm for MAPPO to solve Markov games with time-varying constraints. Particularly, the principal component for the surrogate loss of the conventional MAPPO utilizes both the advantage functions in \eqref{def:am} and the probability ratios in \eqref{def:rm}. The former ones are directly derived from the value functions, which are estimated based on the generated trajectory in \eqref{def:vm}. Now, with the modification in \eqref{def:bara}, the agent action selected at the states satisfying $b_m(t)>0$ in the generated trajectory is the optimal agent action since no other agent action is allowable at these states. Thus, based on the trajectory generated in this way, the value functions and advantage functions can be estimated faster and more accurately. The latter ones, i.e., the probability ratios, are modified by \eqref{def:rm} and the reason for this modification is quite straightforward: when encountering the states satisfying $b_m(t)>0$, the modified actor network by \eqref{def:bara} always selects $J+2$ as the action. Thus, the probability ratio at these states is equal to $\frac{1}{1}=1$.
\end{itemize}
\end{Rem}

\subsection{Online applying}
The online applying algorithm is very similar to the offline one while only uses the trained actor networks for PPOs. Moreover, the values of $\bm{x}(t+1)$ and $\bm{G}(t+1)$ can only be derived from the online interactions with the real environment. Thus, we omit the details.

\section{Numerical Results}\label{Section4}
This section evaluates the performance of the proposed algorithm and compares it with the stat-of-the-art AoI-based algorithms. Specifically, we consider a IoT system with $M\!=\!10$ channels collecting data from $I=90$ monitoring devices and $J=10$ traditional devices, where the corresponding action space has a magnitude of $101^{10}$. Each monitoring device monitors one random process, where each random process has $|\mathcal{X}_i|=10$ states. Meanwhile, the self-transition probabilities $\{p_i\}_{i=1}^{90}$ of these random processes satisfy $p_i=0.6,\ 1\leq i\leq 60$ and $p_i=0.9,\ 61\leq i\leq 90$. The consumed time duration $T_j$ for each data transmission of the $j$\ts{th} traditional device is uniformly picked from $\mathcal{T}\triangleq\{1,2,\cdots,10\}$, i.e., $\bm{T}\triangleq[T_1,T_2,\cdots,T_J]^T\in\mathcal{T}^{J\times1}$. The channel gain model refers \cite{gain_model}, where each channel gain $g_{j,m}(t)$ takes value in $\{\bar{g}_{j,m},\bar{g}_{j,m}+1,\cdots,\bar{g}_{j,m}+9\}$, $\bar{g}_{j,m}$ is uniformly picked from $\{0,1,\cdots,40\}$, and $g_{j,m}(t)$ transits to the current value with probability $0.6$ and to two adjacent values with equal probability $0.2$. The bandwidths for all channels are set as $W_m=1$ and the importance weights for all devices $\{w_i\}_{i=1}^{I+J}$ are uniformly picked from $\{1,2\}$. Other parameters are set as $P/N=1$, $\Delta c=0.1$, $C_L=0.1$, $C_U=4000$, $\Delta C=0.1$, $N_B=4000$, $N_U=80$, $\epsilon=0.2$, $c_1=0.5$, and $c_2=0.01$. Moreover, both the actors and critics in WI-MAPPO utilize two hidden layers, each of which has 128 nodes. We compare the proposed WI-MAPPO with two AoI-based algorithms. The first one is AoI-based WI-MAPPO, which utilizes the same structure with WI-MAPPO, while the employed WIAC module is designed for minimizing AoI according to the method in \cite{AoI_04}; The other algorithm, namely age-aware policy (AAP), utilizes the Lyapunov drift optimization and is currently the state-of-the-art algorithm for joint schedules \cite{joint_3}. Remarkably, AAP cannot address the time-varying constraints issue and thus can only be applied in the scenario where the data transmissions for all traditional devices consume only $1$ time slot, i.e., the scenario with $\bm{T}=1^{J\times1}$, where $1^{J\times1}$ is the $J$-by-$1$ vector with all entries as $1$.
\begin{figure}[!tb]
\centering
\begin{minipage}[t]{0.47\textwidth}
\centering
\includegraphics[width=3.1in]{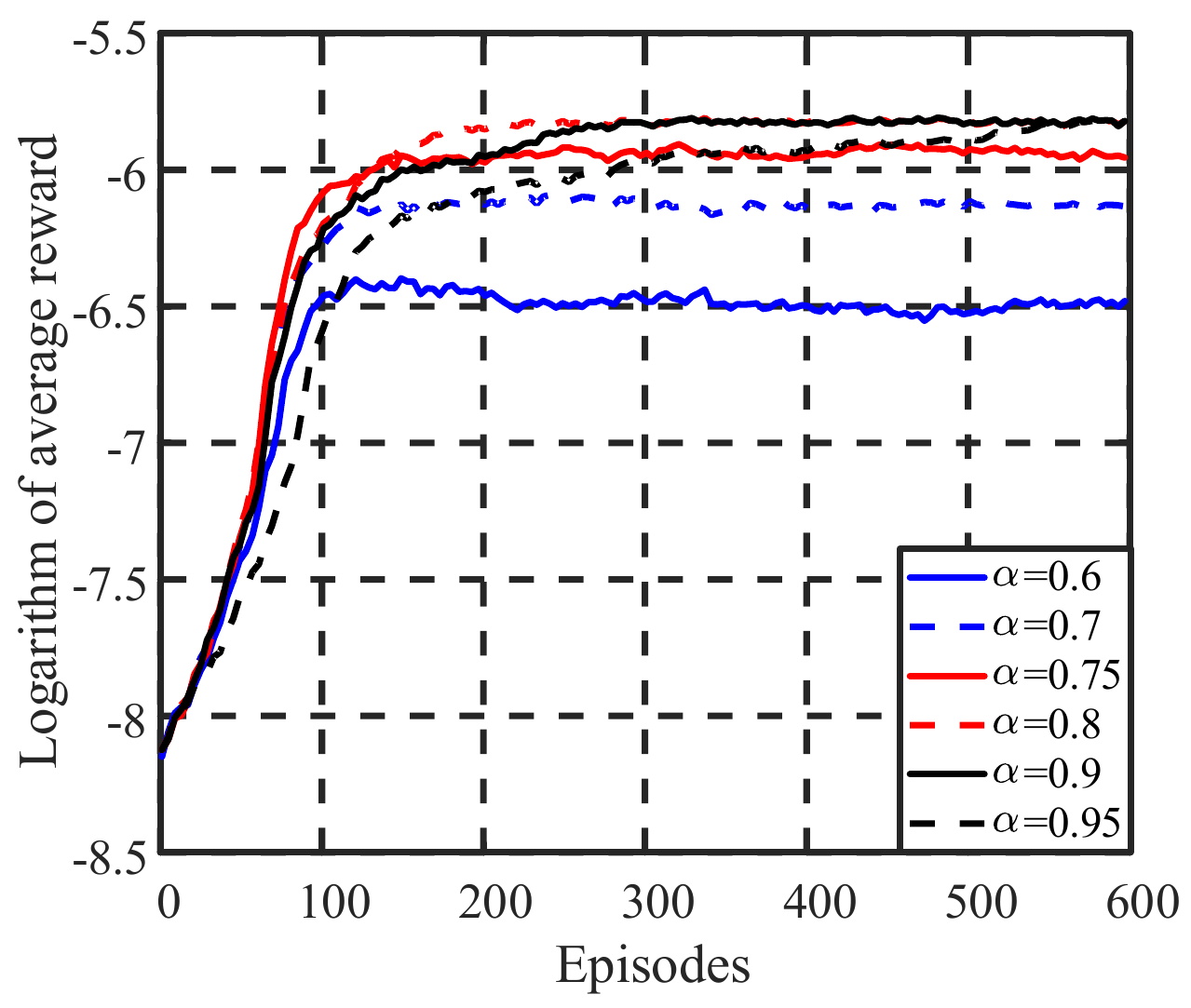}
\caption{Performance comparisons between WI-MAPPOs with different discount factors.}\label{plot_1_1}
\end{minipage}
\begin{minipage}[t]{0.47\textwidth}
\centering
\includegraphics[width=3.1in]{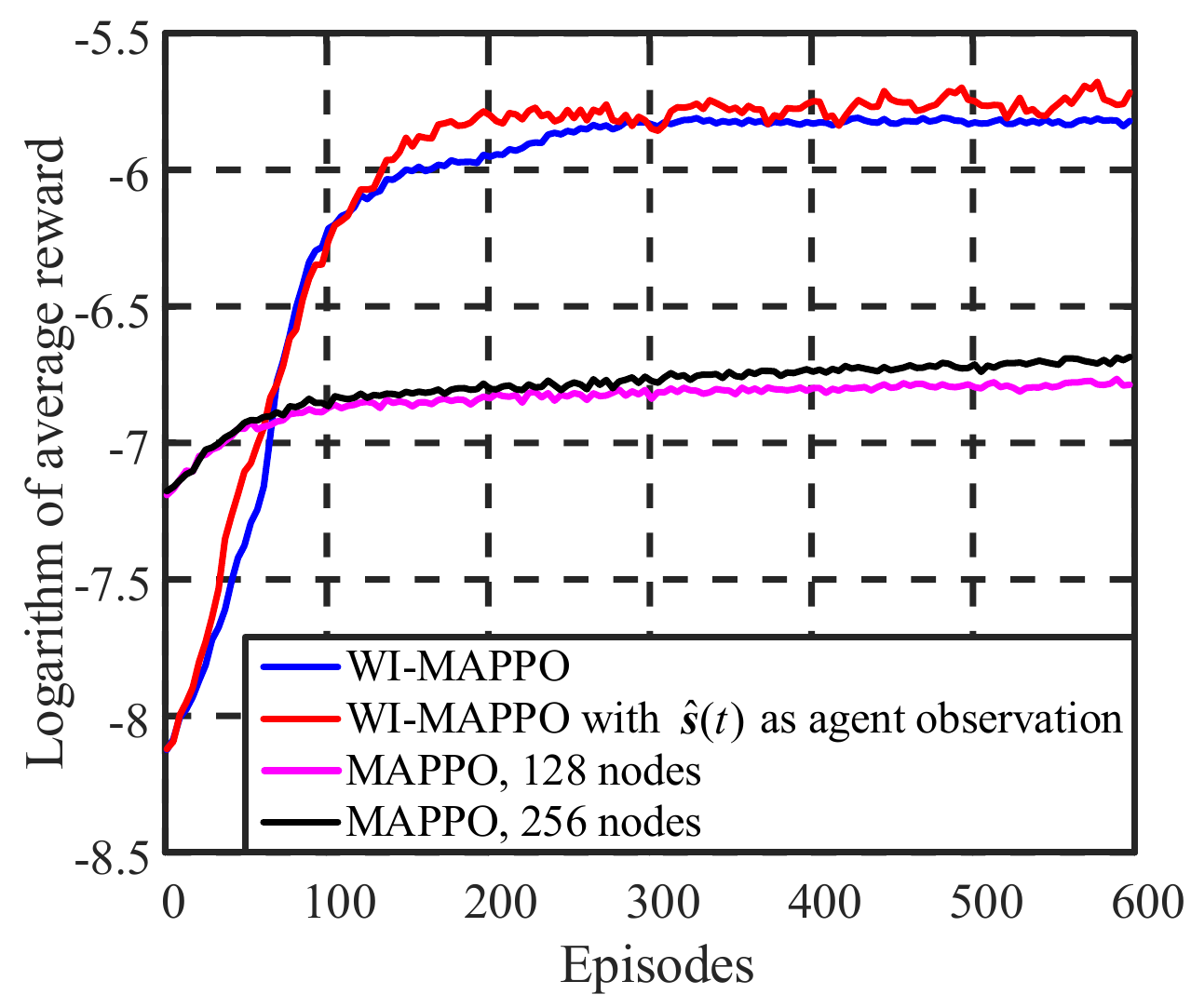}
\caption{Performance comparisons between WI-MAPPO and other MARL algorithms.}\label{plot_1_2}
\end{minipage}
\end{figure}

In Fig.~\ref{plot_1_1}, we investigate the performance of the proposed WI-MAPPO with different discount factors and approximate the value range of the discount factor satisfying the statement in Proposition \ref{prop3}. Specifically, it is observed that when $\alpha$ is no smaller than $0.8$, WI-MAPPO achieves the same maximum on the average reward. This indicates that $[0.8,1)$ could be a proper value range as aforementioned. Meanwhile, it is observed that a too large discount factor, e.g., $\alpha=0.95$, would slow down the convergence. Therefore, we select $\alpha$ as $0.9$ for all the following experiences.

\begin{figure*}[!tb]
\centering
\includegraphics[width=5.8in]{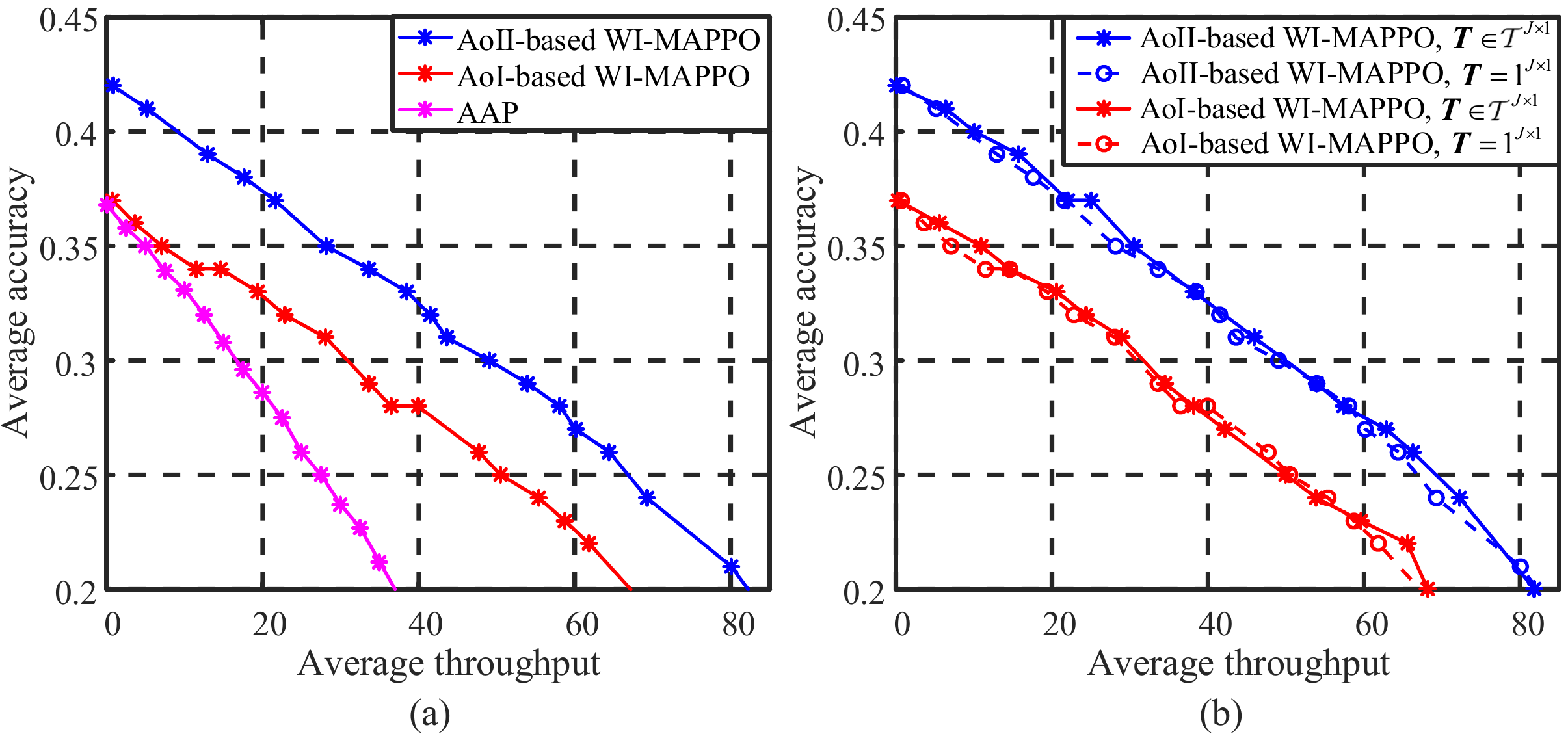}
\caption{Average throughput and average accuracy tradeoff for WI-MAPPOs and AAP in the scenario with 10 channels and 100 devices: (a) tradeoff comparisons between WI-MAPPOs and AAP in the non-constrained case; (b) tradeoff comparisons between WI-MAPPOs in the non-constrained and constrained cases.}\label{plot_2}
\end{figure*}
\begin{figure*}[!htb]
\centering
\includegraphics[width=6in]{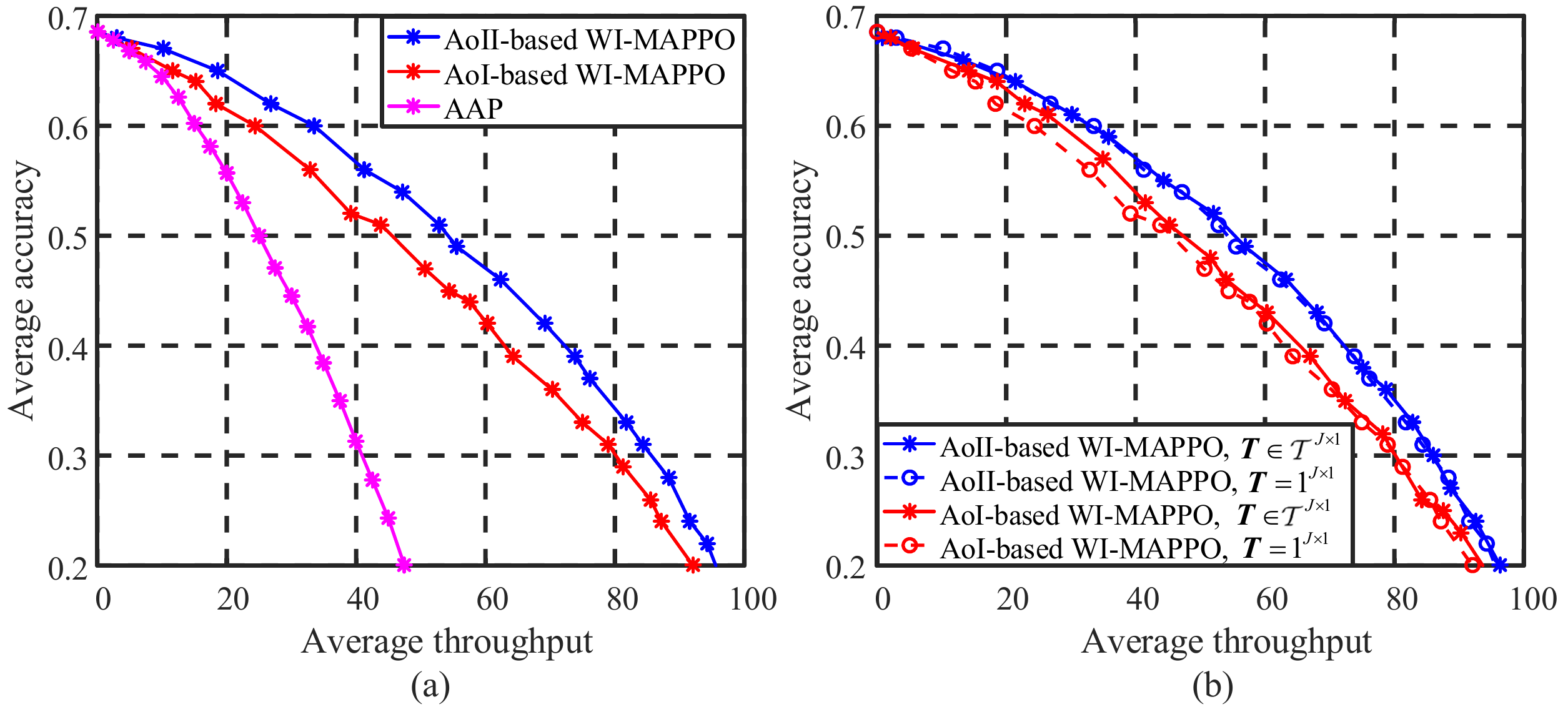}
\caption{Average throughput and average accuracy tradeoff for WI-MAPPOs and AAP in the scenario with 10 channels and 40 devices: (a) tradeoff comparisons between WI-MAPPOs and AAP in the non-constrained case; (b) tradeoff comparisons between WI-MAPPOs in the non-constrained and constrained cases.}\label{plot_3}
\end{figure*}

In Fig.~\ref{plot_1_2}, we validate the performance advantages of the WI-MAPPO over other MARL algorithms. The first algorithm we concerned is the conventional MAPPO. Compared with WI-MAPPO, MAPPO does not have the action space shrinkage provided by the WIAC module and thus suffers from a much larger output dimensionality on its actors, which equals $101$. We test two MAPPO algorithms with separately 128 and 256 nodes on their hidden layers, and it is observed that both of them have far too slow convergence speeds, which also validates Remark \ref{rem31}. The second algorithm simply extends the agent observation from $\bm{s}_m(t)$ to $\hat{\bm{s}}(t)$ and then designs a new WI-MAPPO accordingly. Certainly, this modified WI-MAPPO contains larger PPO actor networks and consumes more offline training and online applying computational resources than the original WI-MAPPO. Moreover, it is observed that it has bare extra gain over the original WI-MAPPO, which also validates Remark \ref{rem21}.

In Fig.~\ref{plot_2}(a), we simulate the scenario satisfying $\bm{T}=1^{T\times1}$, and compare the performances of WI-MAPPO, AoI-based WI-MAPPO, and AAP. It is observed that when the throughput of the traditional devices is not important, the two AoI-based algorithms achieve almost the same average accuracy on predicting the monitored processes, which is around 0.37, and the original WI-MAPPO, which is AoII-based, performs much better and is around 0.42. This validates the advantage of AoII over AoI in pure task-oriented communications. Meanwhile, if we concern the throughput more, both AoII-based and AoI-based WI-MAPPOs gain much larger throughput than AAP when their achieved average accuracies are equal. And AoII-based WI-MAPPO greatly outperforms AoI-based WI-MAPPO. The reason for such performance advantages is also straightforward: AoI captures the aging of sampled status updates, while AoII further factors the semantic of status updates which refers to the usefulness of the transmitted updates relative to prediction of real-time status at data source. Therefore, AoII-oriented scheduling reduces the required data traffic for monitoring devices to guarantee a certain level of prediction performance. In Fig.~\ref{plot_2}(b), we validate the ability of WI-MAPPO on handling the time-varying constraints by comparing its performances in the non-constrained case, i.e., $\bm{T}=1^{J\times1}$, and the constrained case, i.e., $\bm{T}\in\mathcal{T}^{J\times1}$. Remarkably, in the constrained case, the solution space is much smaller than the non-constrained case. Consequently, the optimal performance that any algorithm can achieve in the constrained case is also supposed to be worse than the non-constrained case. However, it is observed that both AoII-based and AoI-based WI-MAPPOs have almost the same performances in these two cases. This amazing result indicates that WI-MAPPO perfectly addresses the time-varying constraints issue for the considered Markov games.

In Fig.~\ref{plot_3}, we simulate a simple scenario with 30 monitoring devices and 10 traditional devices. Since there are only 40 devices requesting for data transmissions, both AoII-based and AoI-based algorithms could achieve the maximum average accuracy, which is around 0.685, when the throughput is not important. Moreover, WI-MAPPO algorithms again greatly outperform AAP. Remarkably, it is observed that AoII-based WI-MAPPO slightly outperforms the AoI-based WI-MAPPO in both the non-constrained and constrained cases. This indicates that the advantages of AoII-based algorithms over AoI-based ones would be more significant when the channel resources are not sufficient.

\section{Conclusions}\label{Section5}
In this work, we study the joint schedule of task-oriented and data-oriented communications and formulate this problem as a challenging Markov decision problem. Insightful techniques and innovative algorithm are utilized to solve this problem as efficiently as possible. Specifically, to simplify this problem, we analysis its ``stationary" feature and Blackwell policies and redesign the reward function based on the channel statistics, by which the solution space and state space are greatly shrunk in an equivalent manner and the optimality criteria is equivalently replaced by a discounted one. To overcome the large discrete action space issue, we convert this problem to an equivalent Markov decision game, where the original action for Markov decision problem is decomposed into low-dimension agent actions. Then, we validate the existence of Whittle's index and design a Whittle's index guided module to further shrink the action space. To overcome the time-varying action constraints issue, we modify the advantage function estimation kernel for MAPPO and extend the training algorithm to solve the constrained Markov games.

\appendices
\setcounter{section}{0}
\section{Proof of Proposition \ref{prop1}}\label{proof1}
It can be easily checked that: problem {\bf (P1)} has infinite and countable states; and the reward of problem {\bf (P1)} satisfies
\begin{align}\label{eq:prop1_up}
    r(t)\leq\sum_{j=1}^Jw_{I+j}\sum_{m=1}^MW_m\log\left(1+\frac{g_{|\mathcal{G}|}P}{N}\right)
\end{align}
and thus it is upper bounded. Based on the above two properties, problem {\bf (P1)} has stationary optimal policies if the following two conditions are satisfied \cite[Proposition 5 \& Theorem 1]{det_sta}:

{\it 1)} Problem {\bf (P1)} has a stationary policy which induces an ergodic Markov chain and has a finite average reward.

{\it 2)} Define $V_{\alpha}(\bm{s})\triangleq\sup_{\boldsymbol{\theta}}\lim_{T\rightarrow\infty}\mathbb{E}_{\pi_{\boldsymbol{\theta}},\text{Pr}\{\bm{s}'|\bm{s},\bm{a}\}}\left[\sum_{t=1}^T\right.$ $\left.\alpha^{t-1}r(t)\Big|_{\bm{s}(1)=\bm{s}}\right]$, where $\alpha\in(0,1)$ is a discount factor; $\boldsymbol{\theta}$ parameterizes policy $\pi_{\boldsymbol{\theta}}$; the expectation is taken with respect to policy $\pi_{\boldsymbol{\theta}}$ and state $\bm{s}(t)$; $V_{\alpha}(\bm{s})$ is the maximum total discounted reward that can be achieved by any policy starting from state $\bm{s}$. Then, there exists a non-negative real number $C$ such that $V_{\alpha}(\bm{s})-V_{\alpha}(\bm{s}_0)\leq C$ holds for all state $\bm{s}\in\mathcal{S}$ and all $\alpha\in(0,1)$, where $\bm{s}_0$ is a reference state in $\mathcal{S}$.

To validate the first condition, we investigate the do-nothing policy $\pi$, where no device would be selected for data transmission over any channel or at any time slot. Apparently, this policy is stationary. In the following, we first show that the average reward under policy $\pi$ is finite, and then show that the induced Markov chain is ergodic.

As for the average reward under policy $\pi$, it equals $\liminf_{T\rightarrow\infty}\mathbb{E}_{\pi,\text{Pr}\{\bm{x}'|\bm{x},\bm{a}\}}\left[\frac{1}{T}\sum_{t=1}^T\left(-\sum_{i=1}^Iw_ix_i(t)\right)\right]$. Specifically, the process $\{x_i(t)\}$ under policy $\pi$ forms an ergodic Markov chain, the transition of which is specified by equations \eqref{trans:x2}, \eqref{trans:x3}. And accordingly, the transition equations are $\mu_1=(1-p_i)\mu_0$ and $\mu_x=(1-q_i)\mu_{x-1},\ x=2,3,\cdots,$
where $\mu_x$ is the state occurrence probability of the state $x$. Then, we can solve that $\mu_0=\frac{q_i}{1+q_i-p_i}$ and $\mu_x=\frac{q_i(1-p_i)(1-q_i)^{x-1}}{1+q_i-p_i}, \forall x\in\mathbb{Z}^{+}$. And it follows
\begin{align*}
	&\liminf_{T\rightarrow\infty}\mathbb{E}_{\pi,\text{Pr}\{\bm{x}'|\bm{x},\bm{a}\}}\left[\frac{1}{T}\sum_{t=1}^T\left(-\sum_{i=1}^Iw_ix_i(t)\right)\right]\\
	=&-\sum_{i=1}^Iw_i\sum_{x=1}^{\infty}x\mu_x=-\sum_{i=1}^Iw_i\frac{1-p_i}{(1+q_i-p_i)q_i}<\infty,
\end{align*}
which implies that the average reward under policy $\pi$ is finite.

Now, we show that the process $\{(\bm{x}(t),\bm{G}(t),\bm{B}(t))\}$ under policy $\pi$ induces an ergodic Markov chain. This is pretty obvious, since $\{\bm{B}(t)\}=\{\bm{0}_{J\times M}\}$ holds and processes $\{x_1(t)\},\cdots,\{x_I(t)\}$, $\{g_{1,1}(t)\},\cdots,\{g_{J,M}(t)\}$ are ergodic and independent.
 
To validate the second condition, we show that for all $\alpha\in(0,1)$, $V_{\alpha}(\bm{s})$ is non-increasing with respect to $x_i$ and $b_{j,m}$, where $\bm{s}=(\bm{x},\bm{G},\bm{B})$ is any state in $\mathcal{S}$ and $x_i$ and $b_{j,m}$ are the $i$\ts{th} and $(j,m)$\ts{th} entries of $\bm{x}$ and $\bm{B}$, respectively. Consequently, by selecting $\bm{s}_0$ as $\left(\bm{0}_{I\times 1},\bm{G}_0,\bm{0}_{J\times M}\right)$ with $\bm{G}_0\triangleq\argmax_{\bm{G}\in\mathcal{G}^{J\times M}}V_{\alpha}\left(\left(\bm{0}_{I\times 1},\bm{G},\bm{0}_{J\times M}\right)\right)$, $V_{\alpha}(\bm{s})-V_{\alpha}(\bm{s}_0)\leq 0$ holds for all $\bm{s}\in\mathcal{S}$ and thus the second condition is verified. 

Here, we only prove that $V_{\alpha}(\bm{s})$ is non-increasing with respect to $x_i$, and the proof for $b_{j,m}$ is very similar and thus omitted. First of all, based on the upper bound on reward in inequality \eqref{eq:prop1_up}, we compute the upper bound of $V_{\alpha}(\bm{s})$, denoted as $V_{\alpha,0}(\bm{s})$, as
\begin{align*}
    V_{\alpha,0}(\bm{s})&\triangleq\lim_{T\rightarrow\infty}\sum_{t=1}^T\alpha^{t-1}\sum_{j=1}^Jw_{I+j}\sum_{m=1}^MW_m\log\left(1+\frac{g_{|\mathcal{G}|}P}{N}\right)\\
    &=\frac{1}{1-\alpha}\sum_{j=1}^Jw_{I+j}\sum_{m=1}^MW_m\log\left(1+\frac{g_{|\mathcal{G}|}P}{N}\right),
\end{align*} 
which is a constant and independent of $\bm{s}$. Next, define $V_{\alpha,n}(\bm{s})$ for $n\in\mathbb{Z}^{+}$ by
\begin{align*}
    V_{\alpha,n}(\bm{s})\triangleq&\max_{\bm{a}\in\mathcal{A}_{\bm{s}}}\Big\{\alpha\sum_{\bm{s}'}\text{Pr}\{\bm{s}'|\bm{s},\bm{a}\}V_{\alpha,n-1}(\bm{s}')\\
    &+r(t)|_{\bm{s}(t)=\bm{s},\bm{a}(t)=\bm{a}}\Big\}.
\end{align*}
Then, we show that $V_{\alpha,n}(\bm{s})$ is non-increasing with respect to $x_i$ for all $n\in\mathbb{Z}_{\geq0}$ by induction:
\begin{itemize}
	\item Rewrite $\bm{s}$ as $(x_i,\bm{x}_{-i},\bm{G},\bm{B})$, where $\bm{x}_{-i}$ consists of all entries of $\bm{x}$ except $x_i$, and construct $\bm{s}^*$ by $\bm{s}^*\triangleq(x_i+1,\bm{x}_{-i},\bm{G},\bm{B})$. Suppose $V_{\alpha,n}(\bm{s})$ is non-increasing with respect to $x_i$, i.e., $V_{\alpha,n}(\bm{s})\geq V_{\alpha,n}(\bm{s}^*)$ holds, which is certainly true for $n=0$.
	\item It can be easily verified that $\mathcal{A}_{\bm{s}}=\mathcal{A}_{\bm{s}^*}$ holds and
    \begin{align*}
    &\alpha\sum_{\bm{s}'}\text{Pr}\{\bm{s}'|\bm{s},\bm{a}\}V_{\alpha,n}(\bm{s}')+r(t)|_{\bm{s}(t)=\bm{s},\bm{a}(t)=\bm{a}}\\
    \geq&\alpha\sum_{\bm{s}'}\text{Pr}\{\bm{s}'|\bm{s}^*,\bm{a}\}V_{\alpha,n}(\bm{s}')+r(t)|_{\bm{s}(t)=\bm{s}^*,\bm{a}(t)=\bm{a}}	
    \end{align*} holds for all $\bm{a}\in\mathcal{A}_{\bm{s}}$. Thus, $V_{\alpha,n+1}(\bm{s})\geq V_{\alpha,n+1}(\bm{s}^*)$ holds.
\end{itemize}
Finally, since for all $\alpha\in(0,1)$, $V_{\alpha,n}(\bm{s})$ converges to $V_{\alpha}(\bm{s})$ as $n$ goes to infinity \cite[Proposition 7.3.1]{dp1}, we know that $V_{\alpha}(\bm{s})\geq V_{\alpha}(\bm{s}^*)$, i.e., $V_{\alpha}(\bm{s})$ is non-increasing with respect to $x_i$.

\section{Proof of Proposition \ref{prop2}}\label{proof2}
To begin with, we rephrase the optimal value of problem {\bf (P1)} as
\begin{align}
    &\max_{\{\bm{a}(t)\}}\ \liminf_{T\rightarrow\infty}\mathbb{E}_{\text{Pr}\{\bm{x}'|\bm{x},\bm{a}\},\text{Pr}\{\bm{G}'|\bm{G}\}}\left[\frac{1}{T}\sum_{t=1}^Tr(t)\right]\nonumber\\
    \begin{split}\nonumber
    	\overset{(i)}=&\max_{\pi\in\Pi^{S}}\lim_{T\rightarrow\infty}\mathbb{E}_{\pi,\text{Pr}\{\bm{x}'|\bm{x},\bm{a}\},\text{Pr}\{\bm{G}'|\bm{G}\}}\left[\frac{1}{T}\sum_{t=1}^T\left(-\sum_{i=1}^Iw_ix_i(t)\right.\right.\\
        &\!\!\left.\left.+\sum_{j=1}^Jw_{I+j}\sum_{m=1}^M\left(\mathbbm{1}_{I+j}\left(a_m(t)\right)+\mathcal{I}\left(b_{j,m}(t)\right)\right)u_{j,m}(t)\right)\right]
    \end{split}\\
    \begin{split}\label{eq:prop2_2}
    	\overset{(ii)}=&\max_{\pi\in\Pi^{S}}\lim_{T\rightarrow\infty}\mathbb{E}_{\pi,\text{Pr}\{\bm{x}'|\bm{x},\bm{a}\},\text{Pr}\{\bm{G}'|\bm{G}\}}\left[\frac{1}{T}\sum_{t=1}^T\left(-\sum_{i=1}^Iw_ix_i(t)\right.\right.\\
        &\left.\left.+\sum_{j=1}^Jw_{I+j}\sum_{m=1}^M\mathbbm{1}_{I+j}\left(a_m(t)\right)\sum_{\tau=t}^{t+T_j-1}u_{j,m}(\tau)\right)\right],
    \end{split}   
\end{align}
where equality $(i)$ holds due to Proposition \ref{prop1} and equality $(ii)$ can be easily derived by combining equalities in \eqref{con:P} and \eqref{trans:b}. In the following, we first show that for each stationary policy $\pi$ for problem {\bf (P1)}, there exists another stationary policy $\hat{\pi}$ for problem {\bf (P2)} such that their objective functions are equal, i.e., \eqref{eq:prop2_2}$|_{\pi}=$\eqref{P2}$|_{\hat{\pi}}$. Then, we show that the inverse holds, too. Based on these two results, problems {\bf (P1)} and {\bf (P2)} are obviously equivalent. 

{\it 1)} We first derive \eqref{eq:prop2_2}$|_{\pi}$. Then, we develop another policy $\bar{\pi}$ for problem {\bf (P1)}, which is simpler than while equivalent to policy $\pi$, i.e., \eqref{eq:prop2_2}$|_{\pi}$=\eqref{eq:prop2_2}$|_{\bar{\pi}}$. Finally, we introduce policy $\hat{\pi}$ for problem {\bf (P2)} and show that \eqref{eq:prop2_2}$|_{\bar{\pi}}=$\eqref{P2}$|_{\hat{\pi}}$. 

To derive \eqref{eq:prop2_2}$|_{\pi}$, we list the countable states of problem {\bf (P1)} as $\bm{s}_1$, $\bm{s}_2$, $\cdots$, $\bm{s}_l$, $\cdots$, respectively, where $\bm{s}_l=(\bm{x}_l,\bm{G}_l,\bm{B}_l)$ is regarded as the $l$\ts{th} state in $\mathcal{S}$ and the $i$\ts{th} entries of $\bm{x}_l$ is denoted as $x_{l,i}$. Then, denote the state occurrence distribution for problem {\bf (P1)} under stationary policy $\pi$ as $\boldsymbol{\mu}^{\pi}\triangleq[\mu_1^{\pi},\mu_2^{\pi},\cdots,\mu_{l}^{\pi},\cdots]^T$, where $\mu_l^{\pi}(\geq0)$ is the state occurrence probability of the state $\bm{s}_l$. And it follows
\begin{align}
    &\eqref{eq:prop2_2}|_{\pi}\nonumber\\
    \begin{split}\nonumber
    	\overset{(i)}=&\sum_{l=1}^{\infty}\mu_l^{\pi}\mathbb{E}_{\bm{a}\sim\pi(\bm{s}_l)}\left[\mathbb{E}_{\text{Pr}\{\bm{x}'|\bm{x},\bm{a}\},\text{Pr}\{\bm{G'}|\bm{G}\}}\left[\left(-\sum_{i=1}^Iw_ix_{l,i}\right.\right.\right.\\
    &\!\!\!\!+\!\!\!\left.\left.\left.\sum_{j=1}^J\!w_{I+j}\!\!\sum_{m=1}^M\!\!\mathbbm{1}_{I+j}\left(a_m\right)\!\!\!\sum_{\tau=0}^{T_j-1}\!\!\!u_{j,m}(\tau)|_{g_{j,m}(0)=[{\bm{G}_l}]_{(j,m)}}\!\!\!\right)\!\!\Bigg|_{\bm{s}_l,\bm{a}}\!\!\right]\!\right]
    \end{split}\\
    \begin{split}
    	=&\sum_{l=1}^{\infty}\mu_l^{\pi}\left(-\sum_{i=1}^Iw_ix_{l,i}\Big|_{\bm{x}_l}+\mathbb{E}_{\bm{a}\sim\pi(\bm{s}_l)}\mathbb{E}_{\text{Pr}\{\bm{G'}|\bm{G}\}}\right.\\
        &\!\!\left.\left[\!\sum_{j=1}^J\!w_{I+j}\!\!\sum_{m=1}^M\!\!\mathbbm{1}_{I+j}\left(a_m\big|_{\bm{a}}\right)\!\!\sum_{\tau=0}^{T_j-1}u_{j,m}(\tau)\Big|_{g_{j,m}(0)=[{\bm{G}_l}]_{(j,m)}}\right]\right)
    \end{split}\nonumber\\
    \begin{split}\label{eq:prop2_4}
    	\overset{(ii)}=&\sum_{l=1}^{\infty}\mu_l^{\pi}\left(-\sum_{i=1}^Iw_ix_{l,i}\Big|_{\bm{x}_l}+\mathbb{E}_{\bm{a}\sim\pi(\bm{s}_l)}\left[\sum_{j=1}^Jw_{I+j}\right.\right.\\
    	&\left.\left.\sum_{m=1}^M\mathbbm{1}_{I+j}\left(a_m\big|_{\bm{a}}\right)\bar{u}_{j,m}(0)\Big|_{g_{j,m}(0)=[{\bm{G}_l}]_{(j,m)}}\right]\right),
    \end{split}    
\end{align}
where $\bm{a}=[a_1,a_2,\cdots,a_M]^T$ is the action in $\mathcal{A}_{\bm{s}_l}$; $\pi(\bm{s}_l)$ is the action distribution at state $\bm{s}_l$ under policy $\pi$; equality $(i)$ holds since policy $\pi$ is stationary; equality $(ii)$ holds due to the definition in equality \eqref{def:baru}. An essential observation is that \eqref{eq:prop2_4} does not involve $\!\bm{B}_l$. Thus, we realign it as
\begin{align}
    &\eqref{eq:prop2_4}\nonumber\\
	\begin{split}\label{eq:prop2_5}
    	\overset{(i)}=&\sum_{\hat{\bm{s}}\in\hat{\mathcal{S}}}\sum_{l\in\{l|\bm{s}_l\in N(\hat{\bm{s}})\}}\mu_l^{\pi}\left(-\sum_{i=1}^Iw_ix_{i}\Big|_{\bm{x}}+\mathbb{E}_{\bm{a}\sim\pi(\bm{s}_l)}\right.\\
        &\!\!\!\!\left.\left[\sum_{j=1}^Jw_{I+j}\sum_{m=1}^M\mathbbm{1}_{I+j}\left(a_m\big|_{\bm{a}}\right)\bar{u}_{j,m}(0)\Big|_{g_{j,m}(0)=[{\bm{G}}]_{(j,m)}}\right]\right)
    \end{split}\\
    \begin{split}\label{eq:prop2_6}
    	\overset{(ii)}=&\!\sum_{\hat{\bm{s}}\in\hat{\mathcal{S}}}\!\mu^{\pi}\!(N(\hat{\bm{s}}))\!\!\left(\!\!-\!\!\sum_{i=1}^I\!w_ix_{i}\Big|_{\bm{x}}\!\!\right)\!\!+\!\!\sum_{\hat{\bm{s}}\in\hat{\mathcal{S}}}\sum_{l\in\{l|\bm{s}_l\in N(\hat{\bm{s}})\}}\mu_l^{\pi}\mathbb{E}_{\bm{a}\sim\pi(\bm{s}_l)}\\
        &\left[\sum_{j=1}^Jw_{I+j}\sum_{m=1}^M\mathbbm{1}_{I+j}\left(a_m\big|_{\bm{a}}\right)\bar{u}_{j,m}(0)\Big|_{g_{j,m}(0)=[{\bm{G}}]_{(j,m)}}\right]
    \end{split}
\end{align}
where in equality $(i)$, $\bm{x}$ and $\bm{G}$ are the components of $\hat{\bm{s}}$, i.e., $\hat{\bm{s}}=(\bm{x},\bm{G},\bm{b})$ holds, and $N(\hat{\bm{s}})$ is the state set defined as $N(\hat{\bm{s}})\triangleq$ $\{\bm{s}_l=(\bm{x}_l,\bm{G}_l,\bm{B}_l)\in\mathcal{S}|\bm{x}_l=\bm{x};\bm{G}_l=\bm{G};\sum_{j=1}^J[\bm{B}_l]_{(j,m)}=b_m, \forall m\in\{1,2,\cdots,M\}\}$ with $b_m$ being the $m$\ts{th} entry of $\bm{b}$; in equality $(ii)$, $\mu^{\pi}(N(\hat{\bm{s}}))$ is the summation of the state occurrence probabilities of all the states in state set $N(\hat{\bm{s}})$ under policy $\pi$ and defined by $\mu^{\pi}(N(\hat{\bm{s}}))\triangleq\sum_{l\in\{l|\bm{s}_l\in N(\hat{\bm{s}})\}}\mu_l^{\pi}$.

Now, we introduce a new policy $\bar{\pi}$ for problem {\bf (P1)} and prove that $\eqref{eq:prop2_6}=\eqref{eq:prop2_2}|_{\bar{\pi}}$ holds. Specifically, denote $\bar{\pi}(\bm{s},\bm{a})$ and $\pi(\bm{s},\bm{a})$ as the probabilities of applying action $\bm{a}$ at state $\bm{s}=(\bm{x},\bm{G},\bm{B})$ under policies $\bar{\pi}$ and $\pi$, respectively. Then, construct policy $\bar{\pi}$ as
\begin{align}\label{eq:prop2_7}
	\bar{\pi}(\bm{s},\bm{a})=\sum_{l\in\{l|\bm{s}_l\in N(\hat{\bm{s}})\}}\mu_l^{\pi}\pi(\bm{s}_l,\bm{a}).
\end{align}
Here, we highlight again that $\hat{\bm{s}}=(\bm{x},\bm{G},\bm{b})$ at the subscript of the RHS of \eqref{eq:prop2_7} is induced from $\bm{s}$ by using $\bm{b}=[b_1,b_2,\cdots,b_M]^T$ and $b_m=\sum_{j=1}^J[\bm{B}]_{(j,m)}$. Obviously, the action distribution that policy $\bar{\pi}$ follows at state $\bm{s}$ is actually the expected action distribution that policy $\pi$ follows over the state set $N(\hat{\bm{s}})$ and the input of policy $\bar{\pi}$ needs to know only $(\hat{\bm{s}}, \bm{a})$ rather than $(\bm{s}, \bm{a})$. Therefore, policy $\bar{\pi}$ is simpler than policy $\pi$ and accordingly we directly denote the action distribution at state $\bm{s}$ under policy $\bar{\pi}$ as $\bar{\pi}(\hat{\bm{s}})$. Based on equality \eqref{eq:prop2_7}, \eqref{eq:prop2_6} is equal to
\begin{align}
	\begin{split}\label{eq:prop2_8}
    	&\sum_{\hat{\bm{s}}\in\hat{\mathcal{S}}}\mu^{\pi}(N(\hat{\bm{s}}))\left(-\sum_{i=1}^Iw_ix_{i}\Big|_{\bm{x}}\right)+\sum_{\hat{\bm{s}}\in\hat{\mathcal{S}}}\mu^{\pi}(N(\hat{\bm{s}}))\mathbb{E}_{\bm{a}\sim\bar{\pi}(\hat{\bm{s}})}\\
        &\left[\sum_{j=1}^Jw_{I+j}\sum_{m=1}^M\mathbbm{1}_{I+j}\left(a_m\big|_{\bm{a}}\right)\bar{u}_{j,m}(0)\Big|_{g_{j,m}(0)=[{\bm{G}}]_{(j,m)}}\right].
    \end{split}
\end{align}
Now, we prove that $\eqref{eq:prop2_8}=\eqref{eq:prop2_2}|_{\bar{\pi}}$ holds, and an essential step is to show that $\mu^{\pi}(N(\hat{\bm{s}}))=\mu^{\bar{\pi}}(N(\hat{\bm{s}}))$ holds for all $\hat{\bm{s}}\in\hat{\mathcal{S}}$. Specifically, it can be easily checked that the transition probabilities of the state sets are equivalent in policies $\pi$ and $\bar{\pi}$, i.e., $\text{Pr}_{\pi}\{N(\hat{\bm{s}}')|N(\hat{\bm{s}})\}\triangleq\mathbb{E}_{\bm{a}\sim\pi(N(\hat{\bm{s}}))}[\text{Pr}$ $\{N(\hat{\bm{s}}')|N(\hat{\bm{s}}),\bm{a}\}]$ equals $\text{Pr}_{\bar{\pi}}\{N(\hat{\bm{s}}')|N(\hat{\bm{s}})\}\triangleq\mathbb{E}_{\bm{a}\sim\bar{\pi}(N(\hat{\bm{s}}))}\left[\text{Pr}\{N(\hat{\bm{s}}')|N(\hat{\bm{s}}),\bm{a}\}\right]$ for all $\hat{\bm{s}},\hat{\bm{s}}'\in\hat{\mathcal{S}}$. Consequently, the Markov chains for state sets under policies $\pi$ and $\bar{\pi}$ are the same. And the occurrence probability of each state set is unique and can be derived by solving the Cerso limit of the Markov chain. Therefore, $\mu^{\pi}(N(\hat{\bm{s}}))=\mu^{\bar{\pi}}(N(\hat{\bm{s}}))$ holds for all $\hat{\bm{s}}\in\hat{\mathcal{S}}$. Based on this condition, it follows
\begin{align}
	\begin{split}
    	\eqref{eq:prop2_8}\!=\!&\sum_{\hat{\bm{s}}\in\hat{\mathcal{S}}}\!\mu^{\bar{\pi}}(N(\hat{\bm{s}}))\!\!\left(\!-\!\sum_{i=1}^Iw_ix_{i}\Big|_{\bm{x}}\right)+\sum_{\hat{\bm{s}}\in\hat{\mathcal{S}}}\mu^{\bar{\pi}}(N(\hat{\bm{s}}))\mathbb{E}_{\bm{a}\sim\bar{\pi}(\hat{\bm{s}})}\\
        &\left[\sum_{j=1}^Jw_{I+j}\sum_{m=1}^M\mathbbm{1}_{I+j}\left(a_m\big|_{\bm{a}}\right)\bar{u}_{j,m}(0)\Big|_{g_{j,m}(0)=[{\bm{G}}]_{(j,m)}}\right]
    \end{split}\label{eq:prop2_9}\\
    =&\eqref{eq:prop2_2}|_{\bar{\pi}}.\nonumber
\end{align}

Finally, we construct the policy $\hat{\pi}$ for problem {\bf (P2)} by
\begin{align}\label{eq:prop2_10}
	\hat{\pi}(\hat{\bm{s}},\bm{a})=\sum_{l\in\{l|\bm{s}_l\in N(\hat{\bm{s}})\}}\mu_l^{\pi}\pi(\bm{s}_l,\bm{a}).
\end{align}
and show that $\eqref{eq:prop2_2}|_{\bar{\pi}}=\eqref{P2}|_{\hat{\pi}}$. Specifically, based on \eqref{eq:prop2_7} and \eqref{eq:prop2_10}, the policy $\bar{\pi}$ for problem {\bf (P1)} makes the same decision with the policy $\hat{\pi}$ for problem {\bf (P2)} when their encountering states are $\hat{\bm{s}}$ and $\bm{s}$, respectively. And based on \eqref{eq:prop2_9}, the instant reward for problem {\bf (P1)} under policy $\bar{\pi}$ can be equivalently regarded as
\begin{align*}
	-\sum_{i=1}^Iw_ix_{i}(t)+\sum_{j=1}^Jw_{I+j}\sum_{m=1}^M\mathbbm{1}_{I+j}\left(a_m(t)\right)\bar{u}_{j,m}(t),
\end{align*}
which is exactly the instant reward $\hat{r}(t)$ for problem {\bf (P2)}. Consequently, the average rewards of problem {\bf (P1)} under policy $\bar{\pi}$ is equals to that of problem {\bf (P2)} under policy $\hat{\pi}$, i.e., $\eqref{eq:prop2_2}|_{\bar{\pi}}=\eqref{P2}|_{\hat{\pi}}$ holds.

{\it 2)} As for the proof for the inverse, i.e., for each stationary policy $\hat{\pi}$ for problem {\bf (P2)}, there exists another stationary policy $\pi$ for problem {\bf (P1)} such that $\eqref{eq:prop2_2}|_{\pi}=\eqref{P2}|_{\hat{\pi}}$ holds, the utilized techniques are similar to the proof in {\it 1)} and thus omitted. We only highlight that the policy $\pi$ for problem {\bf (P1)} is constructed by $\pi(\bm{s},\bm{a})=\hat{\pi}(\hat{\bm{s}},\bm{a})$.

\section{Sketch Proof of Proposition \ref{prop3}}\label{proof3}
Define $\hat{V}_{\alpha}(\hat{\bm{s}})$ by $\hat{V}_{\alpha}(\hat{\bm{s}})\triangleq\sup_{\boldsymbol{\theta}}\lim_{T\rightarrow\infty}\mathbb{E}_{\pi_{\boldsymbol{\theta}},\text{Pr}\{\hat{\bm{s}}'|\hat{\bm{s}},\bm{a}\}}$ $\left[\sum_{t=1}^T\alpha^{t-1}\right.$ $\left.\hat{r}(t)\big|_{\hat{\bm{s}}(1)=\hat{\bm{s}}}\right]$, where $\hat{\bm{s}}$ can be any state in $\hat{\mathcal{S}}$; $\alpha\in(0,1)$ is a discount factor; $\boldsymbol{\theta}$ parameterizes the policy $\pi_{\boldsymbol{\theta}}$. Similar to the proof in Appendix \ref{proof1}, we can show that $\hat{V}_{\alpha}(\hat{\bm{s}})$ is non-increasing with respect to $x_i$ and consequently, the optimal policies for problem {\bf (P3)} are of threshold type with respect to $x_i$: if the optimal policy is to transmit data for the $i$\ts{th} monitoring device at state $(x_i,\bm{x}_{-i},\bm{G},\bm{b})$, it also transmits data for the $i$\ts{th} monitoring device at state $(x_i+1,\bm{x}_{-i},\bm{G},\bm{b})$. Since the optimal threshold cannot be infinitely large, there are finite number of threshold-type policies possibly being the optimal policies. Accordingly, based on \cite[Proposition 4.1.3]{dp2}, there exist Blackwell optimal policies for problem {\bf (P3)}, and based on \cite[Proposition 4.1.7]{dp2}, these policies optimize problem {\bf (P2)}. Thus, Proposition \ref{prop3} holds.

\section{Proof of Proposition \ref{prop4}}\label{proof4}
We first introduce the explicit formulation of the decoupled sub-problems. Next, we validate that the optimal policies for these problems are of threshold type. Then, we derive the optimal policies for these sub-problems. Finally, we validate the existence of the Whittle's index for problem {\bf (P5)}.
\subsubsection{Decoupled sub-problem}
To decouple the problem {\bf (P5)}, we let all monitoring devices be selfish such that each of them aims to minimize its own average weighted AoII. Moreover, each monitoring device is allowed to transmit data at any time slot as long as it pays an additional cost $C$ for each transmission. The goal of each sub-problem is to find the optimal scheduling policy which strikes the balance between the average additional costs and the average weighted AoII for each monitoring device. We formulate the $i$\ts{th} sub-problem as the following Markov decision problem.
\begin{itemize}
	\item state: $x_i(t)\in\mathbb{Z}_{\geq0}$;
	\item action: $a(t)\in\{0,1\}$, where $a(t)=1$ means to transmit data for the $i$\ts{th} monitoring device at the $t$\ts{th} time slot and $a(t)=1$ means not;
	\item transitions: $\text{Pr}\{x_i(t+1)=0|a(t)=1\}=p_i$; $\text{Pr}\{x_i(t+1)=x_i(t)+1|a(t)=1\}=1-p_i$; $\text{Pr}\{x_i(t+1)=0|a(t)=0,x_i(t)=0\}=p_i$; $\text{Pr}\{x_i(t+1)=1|a(t)=0,x_i(t)=0\}=1-p_i$; $\text{Pr}\{x_i(t+1)=0|a(t)=0,x_i(t)>0\}=q_i$; $\text{Pr}\{x_i(t+1)=x_i(t)+1|a(t)=0,x_i(t)>0\}=1-q_i$;
	\item cost: $c(t)=x_i(t)+a(t)C$;
	\item optimality criteria: {\it lim} average optimality criteria.
\end{itemize}

\subsubsection{The structure of the optimal policy}We first study the decoupled sub-problem with total discounted cost criteria and analyze the corresponding optimal policies. Specifically, define $V_{\alpha}(x)$ by $V_{\alpha}(x)\triangleq\inf_{\boldsymbol{\theta}}\lim_{T\rightarrow\infty}\mathbb{E}_{\pi_{\boldsymbol{\theta}},\text{Pr}\{x_i'|x_i,a\}}\left[\sum_{t=1}^T\alpha^{t-1}\right.$ $\left.c(t)\big|_{x_i(1)=x}\right]$, where $\alpha\in(0,1)$ is a discount factor; $\boldsymbol{\theta}$ parameterizes the policy $\pi_{\boldsymbol{\theta}}$. Similar to the proof in Appendix \ref{proof1}, we can show that $V_{\alpha}(x)$ is non-decreasing with respect to $x$, i.e., $V_{\alpha}(x)\leq V_{\alpha}(x+1)$ holds for all $x\in\mathbb{Z}_{\geq0}$. Now, we show that the optimal policy for the discounted version of the decoupled sub-problem is of threshold type: 1) consider the optimal policy is to transmit data for the $i$\ts{th} monitoring device at the state $x\in\mathbb{Z}^{+}$; 2) then, based on Bellman's optimality equation (Prop 7.3.1 in \cite{dp1}), $x+C+\alpha(p_iV_{\alpha}(0)+(1-p_i)V_{\alpha}(x+1))\leq x+\alpha(q_iV_{\alpha}(0)+(1-q_i)V_{\alpha}(x+1))$ holds; 3) since $V_{\alpha}(x+1)\leq V_{\alpha}(x+2)$ holds, $x+1+C+\alpha(p_iV_{\alpha}(0)+(1-p_i)V_{\alpha}(x+2))\leq x+1+\alpha(q_iV_{\alpha}(0)+(1-q_i)V_{\alpha}(x+2))$ holds, too; 4) thus, the optimal policy will also transmit data for the $i$\ts{th} monitoring device at the state $x+1$, which completes the proof.

Similar to the proof in Appendix \ref{proof3}, there are finite number of threshold-type policies possibly being the optimal policies for the discounted decoupled sub-problems. Then, based on Prop 4.1.3 and Prop 4.1.7 in \cite{dp2}, there exist threshold-type optimal policies for the original decoupled sub-problem with {\it lim} average optimality criteria.

\subsubsection{The derivation of the optimal policy}
To derive the optimal policy, we randomly investigate a threshold-type policy $\pi_{x_0}$: if $x\geq x_0$, the $i$\ts{th} monitoring device transmits data; if $x<x_0$, the $i$\ts{th} monitoring device does not transmit data. By solving the transition equations, we derive that
\begin{align}\label{eq:prop4_0}
	\mu_{i,x}\!\!=\!\!\left\{\begin{array}{ll}
	\!\!\!\frac{1}{1+\frac{1-p_i}{q_i}-(1-p_i)\left(\frac{1}{q_i}-\frac{1}{p_i}\right)(1-q_i)^{x_0-1}}&\!\!\!x\!\!=\!\!0\\
	\!\!\!(1-p_i)(1-q_i)^{x-1}\mu_{i,0}&\!\!\!x\!\!=\!\!1,2,\cdots,x_0\\
	\!\!\!(1-p_i)^{x-x_0+1}(1-q_i)^{x_0-1}\mu_{i,0}&\!\!\!x\!\!=\!\!x_0\!+\!1,x_0\!+\!2,\cdots,
    \end{array}\right.
\end{align}
where $\mu_{i,x}$ is the state occurrence probability of the state $x$ in the $i$\ts{th} sub-problem under policy $\pi_{x_0}$. And the average cost equals 
\begin{align}
	f_i(x_0, C)\triangleq&\sum_{x=0}^{x_0-1}w_ix\mu_{i,x}+\sum_{x=x_0}^{\infty}(w_ix+C)\mu_{i,x}\nonumber\\
	=&\frac{\beta_1+(\beta_2+\beta_3x_0)(1-q_i)^{x_0}}{\beta_4-\beta_5(1-q_i)^{x_0-1}},\label{eq:prop4_1}
\end{align}
where
\begin{align*}
	&\beta_1=w_i\frac{1-p_i}{q_i^2}>0;\\
	&\beta_2=w_i\frac{(1-p_i)^2}{p_i^2(1-q_i)}-w_i\frac{1-p_i}{q_i^2}+\frac{1-p_i}{p_i(1-q_i)}C;\\
	&\beta_3=w_i\frac{1-p_i}{1-q_i}\left(\frac{1}{p_i}-\frac{1}{q_i}\right)<0;\ \beta_4=1+\frac{1-p_i}{q_i}>0;\\
	&\beta_5=(1-p_i)\left(\frac{1}{q_i}-\frac{1}{p_i}\right)>0.
\end{align*}
Thus, we can derive the optimal policy by finding the optimal threshold $x_i(C)\in\mathbb{Z}^{+}$, which is defined by $x_i(C)=\argmin_{x_0\in\mathbb{Z}^{+}} f_i(x_0,C)$. Notably, it is not easy to derive the exact value of $x_i(C)$ and neither will we derive it. Instead, the analyses here are used to prove the existence of the Whittle's index for problem {\bf (P5)}.

\subsubsection{The indexability for the decoupled sub-problem}
Based on \cite{whittle}, the existence of the Whittle's index is guaranteed if all the sub-problems are indexable. Specifically, we give the explicit definition of the indexability as follows.
\begin{Thm}[indexability]
Define $Z_i(C)=\{x\in\mathbb{Z}_{\geq0}|x<x_i(C)\}$ as the set of states where the optimal policy is not to transmit data for the $i$\ts{th} monitoring device. Then, the $i$\ts{th} decoupled sub-problem is said to be indexable if it follows
\begin{align*}
	C'\geq C\Rightarrow Z_i(C')\supseteq Z_i(C).
\end{align*}
\end{Thm}

Apparently, it is difficult to directly validate the indexability for the decoupled sub-problem based on the above definition. Instead, we refer \cite[Proposition 2.2]{phd}, according to which, the $i$\ts{th} decoupled sub-problem is indexable as long as $\sum_{x=x_0}^{\infty}\mu_{i,x}$ is decreasing with respect to $x_0$. Based on \eqref{eq:prop4_0}, it follows
\begin{align*}
	\sum_{x=x_0}^{\infty}\mu_{i,x}=\frac{1-p_i}{q_i\left(\frac{1+\frac{1-p_i}{q_i}}{(1-q_i)^{x_0-1}}-(1-p_i)(\frac{1}{q_i}-\frac{1}{p_i})\right)},
\end{align*}
and apparently, $\sum_{x=x_0}^{\infty}\mu_{i,x}$ is decreasing with respect to $x_0$. This completes the proof.

\bibliographystyle{IEEEtran}

\end{document}